\pageno=1                                      %FIRST PAGE NUMBER OF CHAPTER
%\input <author's macro filename>                        %AUTHOR'S MACR FILE
%\input ppiv-style                                 %PPIV MACRO/PARAMETER FILE
% ppiv-style.tex
% Macro file for authors of chapter for PPIV
%
%-----------------------------------------------------------------------
% 
% FONT DEFINITIONS:
%
%-----------------------------------------------------------------------
% Basic font definitions borrowed from TeXbook macros
\font\ninerm=cmr9
\font\eightrm=cmr8
\font\sixrm=cmr6
\font\ninei=cmmi9
\font\eighti=cmmi8
\font\sixi=cmmi6
\skewchar\ninei='177 \skewchar\eighti='177 \skewchar\sixi='177
\font\ninesy=cmsy9
\font\eightsy=cmsy8
\font\sixsy=cmsy6
\skewchar\ninesy='60 \skewchar\eightsy='60 \skewchar\sixsy='60

\font\ninebf=cmbx9
\font\eightbf=cmbx8
\font\sixbf=cmbx6
\font\ninett=cmtt9
\font\eighttt=cmtt8
\hyphenchar\tentt=-1 % inhibit hyphenation in typewriter type
\hyphenchar\ninett=-1
\hyphenchar\eighttt=-1
\font\ninesl=cmsl9
\font\eightsl=cmsl8
\font\nineit=cmti9
\font\eightit=cmti8
\newskip\ttglue
\def\tenpoint{\def\rm{\fam0\tenrm}%
  \textfont0=\tenrm \scriptfont0=\sevenrm \scriptscriptfont0=\fiverm
  \textfont1=\teni \scriptfont1=\seveni \scriptscriptfont1=\fivei
  \textfont2=\tensy \scriptfont2=\sevensy \scriptscriptfont2=\fivesy
  \textfont3=\tenex \scriptfont3=\tenex \scriptscriptfont3=\tenex
  \def\it{\fam\itfam\tenit}%
  \textfont\itfam=\tenit
  \def\sl{\fam\slfam\tensl}%
  \textfont\slfam=\tensl
  \def\bf{\fam\bffam\tenbf}%
  \textfont\bffam=\tenbf \scriptfont\bffam=\sevenbf
   \scriptscriptfont\bffam=\fivebf
  \def\tt{\fam\ttfam\tentt}%
  \textfont\ttfam=\tentt
  \tt \ttglue=.5em plus.25em minus.15em
  \normalbaselineskip=12pt
  \let\sc=\eightrm
  \let\big=\tenbig
  \setbox\strutbox=\hbox{\vrule height8.5pt depth3.5pt width0pt}%
  \normalbaselines\rm}
\def\ninepoint{\def\rm{\fam0\ninerm}%
  \textfont0=\ninerm \scriptfont0=\sixrm \scriptscriptfont0=\fiverm
  \textfont1=\ninei \scriptfont1=\sixi \scriptscriptfont1=\fivei
  \textfont2=\ninesy \scriptfont2=\sixsy \scriptscriptfont2=\fivesy
  \textfont3=\tenex \scriptfont3=\tenex \scriptscriptfont3=\tenex
  \def\it{\fam\itfam\nineit}%
  \textfont\itfam=\nineit
  \def\sl{\fam\slfam\ninesl}%
  \textfont\slfam=\ninesl
  \def\bf{\fam\bffam\ninebf}%
  \textfont\bffam=\ninebf \scriptfont\bffam=\sixbf
   \scriptscriptfont\bffam=\fivebf
  \def\tt{\fam\ttfam\ninett}%
  \textfont\ttfam=\ninett
  \tt \ttglue=.5em plus.25em minus.15em
  \normalbaselineskip=10pt % set to 10pt, not standard 11pt of TeX manmacs
  \let\sc=\sevenrm
  \let\big=\ninebig
  \setbox\strutbox=\hbox{\vrule height8pt depth3pt width0pt}%
  \normalbaselines\rm}
\def\eightpoint{\def\rm{\fam0\eightrm}%
  \textfont0=\eightrm \scriptfont0=\sixrm \scriptscriptfont0=\fiverm
  \textfont1=\eighti \scriptfont1=\sixi \scriptscriptfont1=\fivei
  \textfont2=\eightsy \scriptfont2=\sixsy \scriptscriptfont2=\fivesy
  \textfont3=\tenex \scriptfont3=\tenex \scriptscriptfont3=\tenex
  \def\it{\fam\itfam\eightit}%
  \textfont\itfam=\eightit
  \def\sl{\fam\slfam\eightsl}%
  \textfont\slfam=\eightsl
  \def\bf{\fam\bffam\eightbf}%
  \textfont\bffam=\eightbf \scriptfont\bffam=\sixbf
   \scriptscriptfont\bffam=\fivebf
  \def\tt{\fam\ttfam\eighttt}%
  \textfont\ttfam=\eighttt
  \tt \ttglue=.5em plus.25em minus.15em
  \normalbaselineskip=9pt
  \let\sc=\sixrm
  \let\big=\eightbig
  \setbox\strutbox=\hbox{\vrule height7pt depth2pt width0pt}%
  \normalbaselines\rm}
%
% Now in a position to define font sizes for book headers, captions, etc.
\def\headtype{\ninepoint}                 % headers
\def\abstracttype{\ninepoint}             % abstracts
\def\captiontype{\ninepoint}              % figure captions
            % table notes
\def\footnotetype{\ninepoint}             % footnotes
                  % references
\def\refit{\it}                           % italics in references
\font\chaptitle=cmr10 at 11pt             % chapter title font
\rm                                       % make sure we load cmr fonts

%-----------------------------------------------------------------------
%
% GENERAL SPACINGS AND SKIPS:
%
%-----------------------------------------------------------------------
% spacings and spacing definition
\parindent=0.25in                         % paragraph indentation
\parskip=0pt                              % extra skip between paragraphs
\baselineskip=12pt                        % skip between lines
\hsize=4.25truein                         % width of text
\vsize=7.445truein                        % height of text; exactly 45 lines 
\hoffset=1in                              % horizontal offset on page
\voffset=1.0in                           % vertical offset on page

% define skips before and after sections etc.
\newskip\sectionskipamount                % skip before main section
\newskip\aftermainskipamount              %      after  main section
\newskip\subsecskipamount                 %      before subsection
\newskip\firstpageskipamount              %      at top of first page
\newskip\capskipamount                    %      in captions
\newskip\ackskipamount                    %      before acknowledgments
\sectionskipamount=0.2in plus 0.09in
\aftermainskipamount=6pt plus 6pt         % needs to be as much as whole line
\subsecskipamount=0.1in plus 0.04in
\firstpageskipamount=3pc
\capskipamount=0.1in
\ackskipamount=0.15in
\def\sectionskip{\vskip\sectionskipamount}
\def\aftermainskip{\vskip\aftermainskipamount}
\def\subsecskip{\vskip\subsecskipamount} 

\def\capskip{\hskip\capskipamount}

%-----------------------------------------------------------------------
%
% PAGINATION AND HEADING MACROS:
%
%-----------------------------------------------------------------------
% pagination and related macros
\nopagenumbers                            % turn off default TeX numbering
\newcount\firstpageno                     % create count to hold first page no.
\firstpageno=\pageno                      % remember first page no.: pageno
                                          % should have been set as first line
                                          % in chapter before macros are read
\newcount\chapno                          % create count to hold chapter no.

% all pages have a running head, its contents depending on whether
% page is right or left facing. Exceptions are the first page of
% each chapter and pageinserts (not available to authors, only
% for SSSBOOKS use): these have no head
\def\rightheadline{\headtype\phantom{\folio}\hfil\runningtitletext\hfil\folio}
\def\leftheadline{\headtype\folio\hfil\runningauthortext\hfil\phantom{\folio}}
\headline={\ifnum\pageno=\firstpageno\hfil
           \else
              \ifdim\ht\topins=\vsize           % a full pageinsert
                 \ifdim\dp\topins=1sp \hfil     % a rotated full pageinsert
                 \else
                     \ifodd\pageno\rightheadline\else\leftheadline\fi
                 \fi
              \else
                 \ifodd\pageno\rightheadline\else\leftheadline\fi
              \fi
           \fi}

% a footline page number is used for the first page of a chapter only
\def\bottomnumber{\hss\tenrm[\folio]\hss}
\footline={\ifnum\pageno=\firstpageno\bottomnumber\else\hfil\fi}

%-----------------------------------------------------------------------
%
% FORMATS FOR SECTIONS, ABSTRACTS, CAPTIONS, FOOTNOTES, ETC.:
%
%-----------------------------------------------------------------------
% macros to create sections and subsections - note that contrary to good
% TeX practice, the macro parameters are _deliberately_ delimited by
% whitespace, and as a consequence, there must be nothing attached to
% the end of section argument in curly braces. The reason I did it this
% way are obscure, and probably not `good'. It looks like a kludge; it
% probably is. So be it. It works, as long as you follow the rules,
% i.e. use 
%
%  %
%  \mainsection{This is a main section heading}
%  %
%  Here is the text ...
%
\outer\def\mainsection#1
    {\vskip 0pt plus\smallskipamount\sectionskip
     \message{#1}\vbox{\noindent{\bf#1}}\nobreak\aftermainskip\noindent}
 
\outer\def\subsection#1
    {\vskip 0pt plus\smallskipamount\subsecskip
     \message{#1}\vbox{\noindent{\bf#1}}\nobreak\smallskip\nobreak\noindent}
 
% use this when subsection immediately follows mainsection heading
\def\backup{\nobreak\vskip-\baselineskip\nobreak\vskip-\subsecskipamount\nobreak
}

% macros for formatting chapter title, author names, affiliation, abstract, 
% references, figure captions 
\def\title#1{{\chaptitle\leftline{#1}}}
\def\name#1{\leftline{#1}}
\def\affiliation#1{\leftline{\it #1}}
\def\abstract#1{{\abstracttype \noindent #1 \smallskip\vskip .1in}}
\def\ref{\noindent \parshape2 0truein 4.25truein 0.25truein 4truein}
\def\caption{\noindent \captiontype
             \parshape=2 0truein 4.25truein .125truein 4.125truein}

% version of footnote stuff copied from TeXbook format (as opposed to 
% plain TeX version)
\def\footnote#1{\edef\fspafac{\spacefactor\the\spacefactor}#1\fspafac
      \insert\footins\bgroup\footnotetype
      \interlinepenalty100 \let\par=\endgraf
        \leftskip=0pt \rightskip=0pt
        \splittopskip=10pt plus 1pt minus 1pt \floatingpenalty=20000
        \textindent{#1}\bgroup\strut\aftergroup\strut\egroup\let\next}
\skip\footins=12pt plus 2pt minus 4pt % space added when footnote is present
\dimen\footins=30pc % maximum footnotes per page
%
%-----------------------------------------------------------------------
%
% OTHER GENERAL MACROS:
%
%-----------------------------------------------------------------------
% shorthand for \noindent

% macro to force end of sentence space after capital letters with period
\def\@{\spacefactor 1000}

% redefine \, to work in both math and lr mode
\def\,{\pcomma} 
\def\pcomma{\relax\ifmmode\mskip\thinmuskip\else\thinspace\fi}

% define a reasonable version of \simgt and \simlt using \mathpalette
% as explained in the TeXbook p151 and in the definition of \cong in
% Appendix B. Only problem with these is that the vertical spacing
% commands defined in terms of ex do not seem to reflect the actual
% font sizes in \scriptstyle etc. as expected
\def\oversim#1#2{\lower0.5ex\vbox{\baselineskip=0pt\lineskip=0.2ex
     \ialign{$\mathsurround=0pt #1\hfil##\hfil$\crcr#2\crcr\sim\crcr}}}
\def\simgt{\mathrel{\mathpalette\oversim>}}
\def\simlt{\mathrel{\mathpalette\oversim<}}

%RUNNING TITLE AND RUNNING AUTHOR LIST GO HERE
\def\runningtitletext{FROM PRE-STELLAR CORES TO PROTOSTARS}
\def\runningauthortext{Andr\'e, Ward-Thompson, Barsony}
%
% useful symbols:
%\simlt:\la   \simgt:\ga

%\def\lco{\mbox{$L_{\mbox{\tiny CO}}$}}
%
%\def\fco{\mbox{$F_{\mbox{\tiny CO}}$}}
\def\fco{$F_{\rm CO}$~}
\def\fcom{F_{\rm CO} }
\def\lbol{$L_{\rm bol}$~}
\def\lbolm{L_{\rm bol} }
\def\kms{\rm{km\, s^{-1}}}
\def\myr{\rm{M_\odot\,  yr^{-1}}}
\def\cmg{\rm{cm^{2} \, g^{-1}}}
\def\menv{$M_{\rm env}$~}
\def\menvm{M_{\rm env}}
\def\macc{$\dot{M}_{\rm acc}$~}
\def\maccm{\dot{M}_{\rm acc}}
\def\mjet{$\dot{M}_{\rm jet}$~}
\def\mjetm{\dot{M}_{\rm jet}}

\def\vjetm{V_{\rm jet}}
%\def\fent{\mbox{$f_{\mbox{\tiny ent}}$}}

%
%\def\mcore{\mbox{$M_{\mbox{\tiny core}}$}}
%\def\mdisk{\mbox{$M_{\mbox{\tiny disk}}$}}
%
%\def\minf{\mbox{$\dot{M}_{\mbox{\tiny inf}}$}}
%\def\mwin{\mbox{$\dot{M}_{\mbox{\tiny w}}$}}
%\def\mw{\mbox{$\dot{M}_{\mbox{\tiny w}}$}}
%\def\nhd{\mbox{$n_{\mbox{\tiny H}_{2}}$}}
%\def\smm{\mbox{$S_{\mbox{\tiny 1.3mm}}$}}
%\def\sint{\mbox{$S_{\mbox{\tiny 1.3mm}}^{\mbox{\tiny ~int}}$}}
%\def\smmpc{\mbox{$S_{\mbox{\tiny 1.3mm}}^{\mbox{\tiny ~160pc}}$}}
%\def\gcm{\mbox{$g\,cm^{-2}$}}
%

% other useful symbols:
\def\>{$>$}
\def\<{$<$}

\def\ltsima{$\; \buildrel < \over \sim \;$}
\def\simlt{\lower.5ex\hbox{\ltsima}}
\def\gtsima{$\; \buildrel > \over \sim \;$}
\def\simgt{\lower.5ex\hbox{\gtsima}}
\def\arcsec{\hbox{$^{\prime\prime}$}}

\tolerance=10000
%-------------------------------------- Begin Text!

\null
\vskip -3.0pc
\noindent
\abstracttype {To appear in {\it Protostars and Planets IV}, eds. V. Mannings, A.P. Boss, and 
S.S. Russell (Tucson: Univ. of Arizona Press) -- Accepted 1999 January --}
%\firstpageskip
\vskip 3.0pc
\tenpoint
\null

%TITLE GOES HERE
{\baselineskip=14pt
\title{FROM PRE-STELLAR CORES TO PROTOSTARS:}
\title{THE INITIAL CONDITIONS OF STAR FORMATION}
}

%AUTHOR NAMES AND INSTITUTIONS GO HERE
\vskip .3truein
\name{PHILIPPE ANDR\'E}
\affiliation{CEA/DSM/DAPNIA, Service d'Astrophysique, Saclay, France}
\vskip .2truein
\name{DEREK WARD-THOMPSON}
\affiliation{University of Wales, Cardiff, UK}
\vskip .1truein
\leftline{and}
\vskip .1truein
\name{MARY BARSONY}
\affiliation{University of California at Riverside}
\vskip .3truein

%ABSTRACT FOLLOWS HERE
\abstract{The last decade has witnessed significant advances in our observational understanding of the earliest stages of low-mass star formation. 
The advent of sensitive receivers on large radio telescopes such
as the JCMT and IRAM 30m MRT has led to the identification of  
young protostars at the beginning of the main accretion phase 
(`Class~0' objects), 
and has made it possible to probe, for the first time, the inner
density structure of pre-collapse cores. 
Class 0 objects are characterized by strong, centrally-condensed 
dust continuum emission 
at submillimeter wavelengths, very little emission shortward of 
$\sim 10\, \mu$m, 
and powerful jet-like outflows. Direct evidence for gravitational infall 
has been observed toward several of them. They are interpreted as  
accreting protostars which have not yet accumulated the majority of their final
stellar mass. 
In contrast to protostars, pre-stellar cores have flat inner density 
profiles, suggesting the initial conditions for fast 
protostellar collapse depart sometimes significantly from a singular 
isothermal sphere. In the case of non-singular 
initial conditions, the beginning of protostellar evolution is expected 
to feature a brief phase of vigorous accretion/ejection which may  
coincide with Class~0 objects. 
In addition, submillimeter continuum imaging surveys of  
regions of multiple star formation such as Ophiuchus and Serpens 
suggest a picture according to 
which each star in an embedded cluster is 
built from a finite reservoir of mass and the associated IMF  
is primarily determined at the pre-stellar stage of evolution. 
}

%TEXT BEGINS AT THIS POINT

%\mainsection{I.~~PROTOSTARS AND PLANETS IV \hfil \break IN \TeX}
\mainsection{I.~~INTRODUCTION}
\backup
%

%
%\subsection{A.~~Main Conceptual Phases of Low-Mass Star Formation}
%
The formation of low-mass 
%and intermediate-mass ($M_\star \simlt 8\, M_\odot $) 
stars is believed to involve a series  
of conceptually different stages (e.g. Larson 1969, 
Shu, Adams, \& Lizano 1987). 
The first stage corresponds to the fragmentation of a molecular cloud 
into a number of gravitationally-bound cores which are 
initially supported against 
gravity by a combination of thermal, magnetic, and turbulent pressures 
(e.g. Mouschovias 1991, Shu et al. 1987). 
These pre-stellar condensations/fragments 
form and evolve 
%toward higher degrees of central concentration 
as a result of a still poorly understood mechanism, involving    
ambipolar diffusion (e.g. Mouschovias 1991), 
the dissipation of turbulence (e.g. Nakano 1998), 
and/or an outside impulse (e.g. Bonnell et al. 1997). 
%At some point, they become gravitationally unstable 
%and quickly collapse to form a (possibly multiple) star. 
%Once fast collapse sets in, 
Once such a condensation becomes gravitationally unstable and collapses,  
the main {\it theoretical} features of the ensuing dynamical evolution 
have been known since the pioneering work of Larson (1969). 
During a probably brief initial phase, 
the released gravitational energy is freely radiated away and the collapsing fragment stays roughly isothermal. This ``runaway'' isothermal collapse phase   
tends to produce a strong central concentration of matter with a 
radial density gradient approaching $\rho \propto r^{-2}$ at small radii 
essentially independently 
of initial conditions (e.g. Whitworth \& Summers 1985, 
Blottiau et al. 1988, Foster \& Chevalier 1993). 
It ends with the formation of an opaque, hydrostatic protostellar object 
in the center (e.g. Larson 1969, Boss \& Yorke 1995, Bate 1998). 
Numerical simulations in fact predict the successive formations of 
two hydrostatic objects, before and after the dissociation of 
molecular hydrogen respectively (see Boss \& Yorke 1995), but we will not 
distinguish between them here. One then enters the main accretion phase
during which the central object builds up its 
mass ($M_{\star}$) from a surrounding infalling envelope (of mass
$M_{env}$) and accretion disk, while progressively warming up. 
In this chapter, we will refer to the system consisting of the 
central object, plus envelope and disk as an accreting protostar. 
The youngest accreting protostars have $\rm M_{env} >> M_{\star}$,  
and radiate the accretion luminosity 
$L_{acc} \approx GM_{\star}\dot{M}_{acc}/R_{\star}$.  
In the `standard' theory of isolated star formation (Shu et al. 1987, 1993), 
the collapse initial conditions are taken to be (static) 
singular isothermal spheroids 
($\rho \sim (a^2/2\pi G)\, r^{-2}$, cf. Li \& Shu 1996, 1997), 
there is no runaway collapse phase, 
and the accretion rate $\dot{M}_{acc}$ is constant in time  
$\sim a^3/G$, where $a$ is the effective isothermal 
sound speed. With other collapse initial conditions, 
the accretion rate is generally time-dependent (see III--D below).

Observations have shown that the main accretion phase is 
always accompanied by 
a powerful ejection of a small fraction of the accreted material 
in the form of prominent bipolar jets/outflows (e.g. Bachiller 1996). 
These outflows are believed to carry away the excess angular momentum 
of the infalling matter (e.g. K\"onigl \& Pudritz, this volume). 
When the central object 
has accumulated most ($\simgt 90\% $) of its final, main-sequence mass, 
it becomes a pre-main sequence (PMS) star, 
which evolves approximately at fixed mass on the Kelvin-Helmholtz contraction 
timescale (e.g. Stahler \& Walter 1993).
%although accretion of residual amounts of material may still 
%occur through an accretion disk. 
(Note that, during the protostellar accretion phase, stars more massive
than a few $0.1\, M_\odot $ start burning deuterium, while stars with
masses in excess of $\sim 8\, M_\odot $ begin to burn hydrogen  
-- see Palla \& Stahler 1991.)

The details of the earliest stages outlined above are 
still poorly known. Improving our understanding of these early stages 
is of prime importance since to some extent they must govern the origin 
of the stellar initial mass function (IMF).

%Getting at a detailed understanding of the initial stages of star formation  
%is of prime importance if we are to explain the 
%origin of stellar masses. Indeed, the IMF  
%may be largely pre-determined at the pre-stellar stage 
%(see II--E and IV below). Furthermore, it is during the protostellar 
%accretion stage that stars acquire the mass they will have on the zero-age 
%main sequence (ZAMS). 

Observationally, it is by comparing the structure of starless dense cores 
with that of the envelopes surrounding the youngest stellar objects 
%accreting protostars  
that one may hope to estimate the initial conditions for 
protostellar collapse. 
The purpose of this chapter is to review several major advances 
made in this field over the last decade,   
thanks mostly to ground-based (sub)millimeter continuum observations. 
%In the remainder of this section, we summarize the advantages of 
%such observations for the study of the earliest stages of star formation.  
We discuss results obtained on pre-stellar cores and 
young accreting protostars in Sect.~II and Sect.~III, respectively. 
We then combine these two sets of results and conclude in Sect.~IV.

\mainsection{{I}{I}.~~PRE-STELLAR CORES}
\backup
\subsection{A.~~Definition and Identification}

The pre-stellar stage of star formation 
may be defined as the phase in which a gravitationally bound 
core has formed in a molecular cloud, and evolves 
toward higher degrees of central condensation, but 
no central hydrostatic protostellar object exists yet within the core. 

A pioneering survey of isolated dense cores in dark clouds 
was carried out in transitions 
of NH$_3$ by Myers and co-workers (see Myers \& Benson 1983,
%(see Myers et al. 1983, 
Benson \& Myers 1989 and references therein), who catalogued 
about 90 cores. These were separated into starless cores and cores with 
stars (Beichman et al. 1986), on the basis of the presence or absence of an 
embedded source detected by IRAS. The starless NH$_3$ cores were identified by
Beichman et al. as the potential sites of future isolated 
low-mass star formation. 
Other dense core surveys have been 
carried out by Clemens \& Barvainis (1988), 
Wood et al (1994), Bourke et al (1995a,b), Lee \& Myers (1999), and  
Jessop \& Ward-Thompson (1999).
%Subsequently 

Using the 15~m James Clerk Maxwell Telescope (JCMT), 
Ward-Thompson et al. (1994)
observed the 800-$\mu$m dust continuum emission from about 
20 starless NH$_3$ cores from the Benson \& Myers list, mapping 4 of the cores,
and showed that they have larger FWHM sizes than, but comparable 
masses to the envelopes of the youngest 
protostars (Class~0 sources -- see III below). This is consistent with 
starless NH$_3$ cores being pre-stellar in nature 
and the precursors of protostars (see also Mizuno et al. 1994). 
Ward-Thompson et al. also demonstrated that pre-stellar cores do not 
have density profiles which can be modelled by a single scale-free power law,
but instead have flat inner radial density profiles, suggestive of
magnetically-supported cores contracting by ambipolar diffusion 
(see Mouschovias 1991, 1995 and references therein).
Recent molecular line spectroscopy 
of several pre-stellar cores (e.g. Tafalla et al. 1998 and Myers et al.  
this volume) appears 
to support the argument that they are contracting, but more slowly than the 
infall seen toward Class 0 protostars (e.g. Mardones et al. 1997 -- 
see III--C). 

The 800~$\mu$m study by Ward-Thompson et al. (1994) also suggests
that wide-field submillimeter continuum imaging may be a powerful tool to 
search for new pre-stellar cores in the future (cf. Ristorcelli et al. 1998).

%\placefigure{fig1}

\subsection{B.~~Spectral Energy Distributions and Temperatures}

The advent of the Infrared Space Observatory (ISO) of the European Space 
Agency (ESA), has allowed pre-stellar cores to be studied in the far-infrared
for the first time (Ward-Thompson et al. 1999a in prep), 
since these cores were not detected by IRAS. 
%Ward-Thompson et al. (1999 in prep) made
%maps using the photometer on board the ISO spacecraft, ISOPHOT, 
%at 90, 170 and 200$\mu$m. 
Likewise, the Submillimetre Common User
Bolometer Array (SCUBA) camera on the JCMT has allowed pre-stellar cores
to be observed in the submm with greater signal to noise than ever before
(Ward-Thompson et al. 1999b in prep).
Pre-stellar cores emit almost all of their radiation in the FIR/submm/mm
regimes, so the combination of these two instruments provides a unique 
opportunity to study them. 
%For the purposes of this review we will show the
%results for one source, L1544, as illustrative of the results obtained.

As an illustration, 
Figure 1 shows a series of images of the L1544 pre-stellar core at 90 \&
200~$\mu$m (from ISO), 850~$\mu$m (from SCUBA), and 1.3~mm (from IRAM). 
The core is clearly detected at 200--1300~$\mu$m,
but is almost undetected at 90~$\mu$m.  This shows that the core is very cold,
and its dust temperature 
%of the dust which is emitting at these wavelengths 
can be obtained from fitting a modified black-body to the observed 
emission. 
%Figure 2a shows 
The spectral energy distribution (SED) of L1544 in the far-infrared
and submillimeter wavelength regimes is shown in Fig.~2. The solid line
is a grey-body curve of the form: 

\vskip 0.25truecm

$ S_{\nu} = B_{\nu}(T_{dust})\, [1-\rm{exp}(-\tau_\nu)]\, \Delta \Omega \, , $ 

\vskip 0.25truecm

\noindent
where $B_{\nu}(T_{dust})$ is the Planck function at frequency $\nu $ 
for a dust temperature $T_{dust}$, $ \tau_\nu = \kappa_\nu\, \Sigma $ is the
dust optical depth through a mass column density $ \Sigma $ (see below), 
and $ \Delta \Omega $ is the source solid angle. 
In this simple modelling, 
the dust opacity per unit (gas~$+$~dust) mass column density,  
$\kappa_{\nu}$, is  
assumed to scale as $\nu^{\beta}$ with $\beta =$~1.5--2 as usually 
appropriate 
in the submillimeter range (e.g. Hildebrand 1983). 
For L1544, a good fit to the SED is obtained with 
$T_{dust} = 13$~K and $\beta = 2$. Similar results are obtained in other 
starless cores. This confirms the lack of any warm dust 
in such cores, and consequently the lack of any embedded 
protostellar object.  
The (sub)millimeter data show a morphology similar to 
the ISO images, at much higher resolution, indicating that the same dust
is being observed at all wavelengths. Consequently the temperature derived
from the SED is representative of all the emitting dust and  
can be used to convert submillimeter fluxes 
into estimates of dust masses, and hence to gas masses.

%\placefigure{fig2}

\subsection{C.~~Mass and Density Structure}

Dust emission is generally optically thin at (sub)millimeter wavelengths, 
and hence is a direct tracer of the mass content of molecular 
cloud cores.
For an isothermal dust source, the total (gas~$+$~dust) mass  
$M(r<R)$ contained within a radius $R$ from the center  
is related to the submillimeter flux density 
$S_{\nu}(\theta)$ integrated over a circle of 
projected angular radius $\theta = R/d$ by:

\vskip 0.25truecm

$M(r<R)\, \equiv \, \pi R^2 <\Sigma>_R \, = [S_{\nu}(\theta)\, d^2]/[ \kappa_{\nu}\, 
B_{\nu}(T_{dust})]$,  

\vskip 0.25truecm

\noindent
where $<\Sigma>_R$ is the average mass column density. 
The dust opacity $\kappa_{\nu}$ is somewhat uncertain, but the 
uncertainties are much reduced when appropriate dust models are used 
(see Henning et al. 1995 for a review). 
For pre-stellar cores of intermediate densities 
($n_{H_2} \simlt 10^5\ cm^{-3}$), 
$\kappa_{\nu}$ is believed to be close to  
$\kappa_{1.3} = 0.005\ \cmg$ at 1.3~mm
%$\lambda = 1.3$~mm 
(e.g. Hildebrand 1983, Preibisch et al. 1993). 
In denser cloud cores and protostellar envelopes, 
grain coagulation and the formation of ice mantles 
make $\kappa_{\nu}$ a factor of $\sim 2$ larger, i.e., 
$\kappa_{1.3} = 0.01\ \cmg$ assuming a gas-to-dust mass ratio of 100 
%with the same $\beta$ index 
(e.g. Ossenkopf \& Henning 1994). A still higher value, 
$\kappa_{1.3} = 0.02\ \cmg$, is recommended in protoplanetary disks 
(Beckwith et al. 1990, Pollack et al. 1994).

%\noindent
Following this method, FWHM masses ranging from $\sim 0.5\ M_\odot$ 
to $\sim 35\ M_\odot$ are derived for the 9 isolated cores mapped by 
Ward-Thompson, Motte, \& Andr\'e (1999 -- hereafter WMA99).

Although observed pre-stellar cores are generally not circularly or 
elliptically symmetric (see, e.g., Fig.~1), 
%but exhibit some 
%asymmetries at least at the lower contour levels (see, e.g., Fig.~1).  
%Nevertheless, 
one can still usefully constrain  
their radial density profiles by averaging the (sub)millimeter 
emission in circular or elliptical annuli.

%\noindent
Figure 3a shows the azimuthally-averaged radial intensity profile of 
the pre-stellar core L1689B at 1.3~mm, compared to the profile of a
spherical isothermal core model with $\rho(r)\propto r^{-2}$. 
The model intensity profile results from a complete simulation 
taking into account 
both the observing technique (dual-beam mapping) and the reduction method
(cf. Andr\'e, Ward-Thompson, \& Motte 1996).
We see that L1689B exhibits the familiar radial profile  
of pre-stellar cores, with
%which possess 
a flat inner region, steepening toward the edges 
(Ward-Thompson et al. 1994, Andr\'e et al. 1996, WMA99). 
In this representative example, the radial density profile inferred 
assuming a constant dust temperature 
is as flat as $\rho(r) \propto r^{-0.4}$ 
(if the 3-D core shape is disk-like) or $\rho(r) \propto r^{-1.2}$ 
(if the core shape is spheroidal) 
at radii less than $R_{flat} \sim 4000$~AU, and 
approaches $\rho(r) \propto r^{-2}$ only 
between $\sim 4000$~AU and $\sim 15000$~AU.

%\noindent
More recently, it has been possible to constrain the outer density 
gradient of starless cores 
through {\it absorption} studies in the mid-infrared with ISOCAM 
(e.g. Abergel et al. 1996, 1998, Bacmann et al. 1998).
It appears that isolated pre-stellar cores
are often characterized by sharp edges, steeper than $\rho \propto r^{-3}$ 
or $\rho \propto r^{-4}$, at radii $R \simgt 15000$~AU.
%possibly produced by external pressure. 

These features of pre-stellar density structure, i.e., inner flattening and 
sharp outer edge, are qualitatively consistent with models of 
magnetically-supported cores 
evolving through ambipolar diffusion prior to protostar formation 
%undergoing ambipolar diffusion prior to collapse 
(e.g. Ciolek \& Mouschovias 1994, Basu \& Mouschovias 1995), 
although the models generally 
require fairly strong magnetic fields ($\sim 100\ \mu$G).
Alternatively, the observed structure may also be explained by models 
of thermally supported  
self-gravitating cores interacting with 
an external UV radiation field (e.g. Falgarone \& Puget 1985, 
Chi\`eze \& Pineau des For\^ets 1987). 

%\placefigure{fig3}

\subsection{D.~~Lifetimes}
 
Beichman et al. (1986) used the ratio of 
numbers of starless cores to numbers of cores with embedded IRAS sources
to estimate their relative timescales. They found roughly equal numbers
of cores with and without IRAS sources. They estimated the lifetime of
the stage of cores with stars, based on T~Tauri lifetimes and pre-main
sequence HR diagram tracks (e.g. Stahler 1988). Based on this, they
estimated the lifetime of the pre-stellar core phase to be a few times
10$^6$ yr.  

%\noindent
However, the lifetime of a pre-stellar core depends on its 
central density. Figure~4 (taken from Jessop \& Ward-Thompson 1999)
shows the estimated lifetime of starless cores for each of  
six dark cloud surveys mentioned in II--A above, 
versus the mean volume density of cores in each sample. 
The lifespan of cores without stars in each sample was estimated from
the fraction of cores with IRAS sources, using the same method as 
Beichman et al. (1986). 
An anti-correlation between lifetime and density is clearly apparent in Fig.~4.
The solid line has the form $t \propto \rho^{-0.75}$, while the dashed line 
is of the form $t \propto \rho^{-0.5}$. These two forms are expected 
for cores evolving on the 
%supported by a uniform magnetic field and contracting through 
ambipolar diffusion timescale $t_{AD} \propto x_e $ (where $x_e$ is 
the ionisation fraction -- e.g. Nakano 1984), 
if the dominant ionisation mechanism is cosmic ray ionisation 
or UV ionisation, respectively (e.g. McKee 1989).

More details about the evolution of pre-stellar 
cores at higher densities can be inferred from the 
results of the submm continuum SCUBA survey of Ward-Thompson et al. (1999b).
In this survey, 17 of the 38 NH$_3$ 
cores without IRAS sources from Benson \& Myers (1989) 
were detected by SCUBA at 850$\mu$m. 
Ward-Thompson et al. (1999b) estimate that the 17 SCUBA detections 
all have central densities between $\sim$10$^5$cm$^{-3}$ and 
$\sim$10$^6$cm$^{-3}$, whilst the 21 non-detections must have lower central 
densities, typically between $\sim$10$^4$cm$^{-3}$ and $\sim$10$^5$cm$^{-3}$ 
(see Benson \& Myers 1989, Butner et al. 1995, and references therein).  
Consequently, they deduce that the lifetime of these two phases --
central density increasing from $\sim$10$^4$cm$^{-3}$ to
$\sim$10$^5$cm$^{-3}$ compared to central density of
$\sim$10$^5$cm$^{-3}$ until the formation of 
a protostellar object at the center --
must be roughly equal. 
This can be compared with the predictions of ambipolar diffusion models. 

%\noindent
Figure 3b is taken from Ciolek \& Mouschovias (1994) and shows the
radial density profile predicted by an ambipolar diffusion model
at different evolutionary stages ($t_0$ to $t_6$). The 
stage at which the central density is $\sim$10$^4$cm$^{-3}$ corresponds to
time $t_1$ and the stage at which the central density is $\sim$10$^5$cm$^{-3}$
corresponds to time t$_2$. The time at which a protostellar object forms is 
effectively $t_6$. In this model the time taken to go from $t_1$ to $t_2$
($\sim 2 \times 10^6$~yr) is six times longer than the time taken to go 
from $t_2$ to $t_6$.
Some discrepancy could perhaps be accounted for 
by the statistical errors associated with our source number counting technique,
but the ratio between the two timescales should be fairly robust:
The model predicts that SCUBA should only have detected $\sim$1/7 of
the cores, whereas it detected half of the sample.
 
We are left with the conclusion that {\it cores at
central densities of order $\sim$10$^5$cm$^{-3}$ evolve more slowly than 
ambipolar diffusion models predict} -- i.e. the cores experience 
more support than a simple static magnetic field can provide. 
The extra support could perhaps be provided by turbulence  
which generates non-static magnetic fields (e.g. Gammie \& Ostriker 1996, 
Nakano 1998, Balsara et al. 1998).
%Ostriker et al. 1998).
%Vasquez-Semadeni et al. this volume). 

%\placefigure{fig4}

\subsection{E.~~Pre-Stellar Condensations in Star-Forming Clusters}

In regions of multiple star formation, submillimeter dust continuum mapping  
%young, currently forming star clusters 
has revealed a wealth of small-scale cloud fragments, 
sometimes organized along filaments (e.g. Mezger et al. 1992, AWB93, 
Casali et al. 1993, Launhardt et al. 1996, Chini et al. 1997b, 
Johnstone \& Bally 1999). Such fragmentation 
along filaments has not been observed in Taurus, but examples do exist  
in young embedded clusters forming primarily low-mass stars like 
$\rho$~Ophiuchi (Motte, Andr\'e, \& Neri 1998).
The individual fragments, which 
are denser ($<n> \simgt 10^6$--$10^7$~cm$^{-3}$) and 
more compact (a few 1000~AU in size) than the isolated 
pre-stellar cores discussed 
above, often remained totally undetected (in emission) 
by $IRAS$ or ISO in the mid- to far-IR.  
Since molecules tend to freeze out onto dust grains at low temperatures 
and high densities, (sub)millimeter dust emission may be the most 
effective tracer of such condensations (e.g. Mauersberger et al. 1992).  
 
%\noindent
The most centrally-condensed of these starless fragments 
have been claimed to be isothermal protostars, i.e., 
collapsing condensations with no central hydrostatic object   
(see I) (e.g. Mezger et al. 1992, 
Launhardt et al. 1996, Motte et al. 1998). Good examples are 
FIR~3 and FIR~4 in NGC~2024, OphA-SM1 and OphE-MM3 in $\rho$~Oph, LBS~17-SM 
%and LBS~36-SM2 
in Orion~B, or MMS1 and MMS4 in OMC-3. This isothermal 
protostar interpretation remains to be confirmed, however, by observations 
of appropriate spectral line signatures (cf. Myers et al. this volume).

Furthermore, the advent of large-format bolometer arrays 
%on radiotelescopes such as the IRAM~30~m and the JCMT 
now makes possible systematic studies 
of the genetic link between pre-stellar cloud fragments and young stars.
Figure 5 is a 1.3~mm continuum wide-field mosaic 
of the $\rho$~Oph cloud (Motte et al. 1998) 
showing a total of 100 structures with characteristic 
angular scales of $\sim $~15\arcsec--30\arcsec ~(i.e., $\sim $~2000--4000~AU),
%are detected in the mosaic of Fig.~5, 
which are associated with 59 starless condensations (undetected by ISO 
in the mid-IR) and 41 embedded YSOs 
(detected at IR or radio continuum wavelengths). 

%\placefigure{fig5}

Comparison of the masses derived from the 1.3~mm continuum 
(from $\sim$0.05$M_\odot$ to $\sim$3$M_\odot$) 
with Jeans masses suggests that most of the 59 starless fragments 
are close to gravitational virial equilibrium with 
$M/M_{vir} \simgt $~0.3--0.5 and will form stars %(or brown dwarfs) 
in the near future.  
%and hence are pre-stellar in nature.  
These pre-stellar condensations generally have flat inner density 
profiles like isolated pre-stellar cores,
%(see Fig.~4c of Motte et al. 1998), 
but are distinguished by compact, 
finite sizes of a few thousand AU.  
The typical fragmentation lengthscale derived from the average 
projected separation between condensations is $\sim$~6000AU in $\rho$~Oph.
This is $\sim 5$ times smaller than the 
radial extent of isolated dense cores in the Taurus cloud 
(see G\'omez et al. 1993).

Figure~6 shows the mass distribution of the 59 $\rho$~Oph 
pre-stellar fragments. It follows approximately 
$\Delta N/\Delta M  \propto M^{-1.5} $ below $\sim$0.5$M_\odot$, 
which is similar to the clump mass spectrum 
found by large-scale molecular line studies (e.g. Blitz 1993 and 
Williams et al. this volume). 
The novel feature, however, is that the fragment mass spectrum 
found at 1.3~mm in $\rho$~Oph
appears to steepen to $\Delta N/\Delta M  \propto M^{-2.5} $ above  
$\sim$0.5$M_\odot$.   
A similarly steep mass spectrum above $\sim$0.5$M_\odot$ 
was obtained by Testi \& Sargent (1998) 
for compact 3~mm starless condensations in the Serpens core. 
These pre-stellar mass spectra resemble 
%mimic 
the shape   
of the {\it stellar} initial mass function (IMF), which is known to approach 
$\Delta N/\Delta M_\star \propto M_\star^{-2.7}$
for $ 1\, M_\odot \simlt M_\star \simlt 10\, M_\odot$  
and $\Delta N/\Delta M_\star \propto M_\star^{-1.2} $ 
for $ 0.1\, M_\odot \simlt M_\star \simlt 1\, M_\odot$ 
(e.g. Kroupa et al. 1993, Tinney 1993, 1995, Scalo 1998, 
see also Meyer et al. this volume). 
Given the factor of $\sim 2$ uncertainty on the measured pre-stellar masses, 
such a resemblance is remarkable and 
suggests that {\it the IMF of embedded clusters 
is primarily determined at the pre-stellar stage of star 
formation} (see also IV below).

%\placefigure{fig6}

\mainsection{{I}{I}{I}.~~THE YOUNGEST PROTOSTARS}
\backup
\subsection{A.~~Class~0 Protostars and Other YSO Stages}

{\it 1. Infrared YSO classes.}\ \
In the near-/mid-infrared, three broad classes of young stellar objects 
(YSOs) can be distinguished 
based on the slope $\rm \alpha_{IR} = dlog(\lambda
F_\lambda)/dlog(\lambda)$ of their SEDs 
between 2.2~$\mu$m and 10--25~$\mu$m, 
which are interpreted in terms of an evolutionary
sequence (Lada \& Wilking 1984, Lada 1987). 
Going backward in time, Class~III ($\rm \alpha_{IR} < -1.5$)
and Class~II ($\rm -1.5 < \alpha_{IR} < 0$) sources
correspond to PMS stars (``Weak'' and ``Classical'' T~Tauri stars, 
respectively) surrounded by a circumstellar disk (optically thin and
optically thick at $\lambda \simlt 10\ \mu$m, respectively), 
but lacking a dense circumstellar envelope (see Andr\'e 
\& Montmerle 1994 -- hereafter AM94). 
The youngest YSOs detected at 2~$\mu$m are the Class~I sources, 
which are characterized by $\rm \alpha_{IR} > 0$ 
%between $\lambda \sim 2\ \mu$m 
%and  $\lambda \sim 60\ \mu$m 
(e.g. Wilking, Lada, \& Young 1989 -- WLY89), 
and the close association with dense molecular gas (e.g. Myers et al. 1987). 
Class~I objects are now interpreted as relatively evolved protostars 
with typical ages $\sim$~1--2~$\times$~10$^5$~yr (e.g. 
Barsony \& Kenyon 1992, Greene et al. 1994, 
Kenyon \& Hartmann 1995), surrounded by both a disk and a diffuse  
circumstellar envelope of substellar ($\simlt$ 0.1--0.3~$M_\odot$) mass 
(Whitney \& Hartmann 1993, Kenyon et al. 1993b, 
AM94, Lucas \& Roche 1997). Their SEDs are successfully modeled 
in the framework of the ``standard'' theory of isolated protostars 
(e.g. Adams, Lada, \& Shu 1987, Kenyon et al. 1993a), 
in agreement with the idea that they derive 
a substantial fraction of their luminosity from accretion 
(see also Greene \& Lada 1996 and Kenyon et al. 1998). 

%\smallskip
\vskip 0.1truecm
%
%{\it 2.  Class~0 protostars and isothermal protostellar fragments.}\ \
{\it 2.  Class~0 protostars.}\ \ 
Several condensations detected in submillimeter 
dust continuum maps 
of molecular clouds (e.g. II--E) appear to be associated with formed, 
hydrostatic YSOs and have been designated ``Class~0'' protostars 
(Andr\'e, Ward-Thompson, \& Barsony 1993 -- AWB93). 
%The prototype of 
%this class  
%is the $\rho$~Ophiuchi outflow source VLA~1623 (Andr\'e et al. 1990).
Specifically, Class~0 objects   
are defined by the following observational properties (AWB93): 

$\bullet$ (i) Indirect evidence for a central YSO, 
as indicated by, e.g., the detection of a compact centimeter radio continuum 
source, a collimated CO outflow, or an internal heating source.

$\bullet$ (ii) Centrally peaked but extended submillimeter continuum emission 
tracing the presence of a spheroidal circumstellar dust envelope (as opposed 
to just a disk). 

$\bullet$ (iii) High ratio of submillimeter to bolometric luminosity 
suggesting the envelope mass exceeds the central stellar mass: 
L$_{smm}/$L$_{bol} > 0.5$\%, where 
L$_{smm}$ is measured longward of 350~$\mu$m. 
In practice, this often means a SED 
resembling a single temperature blackbody at $\rm T \sim$~15--30~K (see Fig.~2). 
%\oneskip

Property (i) distinguishes Class~0 objects from the pre-stellar 
cores and condensations discussed in Sect.~II. In particular, deep 
VLA observations reveal no compact radio continuum sources in the
centers of pre-stellar cores (Bontemps 1996, Yun et al. 1996).
Properties (ii) and (iii) distinguish Class~0 objects from more evolved 
(Class~I and Class~II) YSOs. As shown by AWB93, 
the L$_{smm}/$L$_{bol}$ ratio should roughly track the ratio 
$M_{env}/M_\star $ of envelope to stellar mass, 
%in low-mass embedded YSOs, 
and may be used as an evolutionary indicator (decreasing with time), 
for low-luminosity ($L_{bol} \simlt 50\, L_\odot $) embedded YSOs. 
Criterion (iii) approximately selects objects which have  
$M_{env}/M_\star > 1$, assuming 
%the most 
plausible relations between L$_{bol}$ and $M_\star$ on the one hand and 
between $\rm L_{smm}$ and $M_{env}$ on the other hand (see AWB93 and AM94).
%for details). 
[A roughly equivalent criterion is   
%S$_{1.3mm}^{int}(d/160pc)^2/$L$_{bol} \sim 0.2$~Jy/L$_\odot$ and 
$M_{env}/L_{bol} > 0.1\ M_\odot/L_\odot$.] 
{\it Class~0 objects are therefore excellent candidates 
for being very young accreting protostars in which a hydrostatic 
core has formed but not yet accumulated the majority of its final mass}. 
%In contrast to Class~I sources, most of the mass is still in the form of a 
%dense circumstellar envelope at the Class~0 stage.
In practice, most of the confirmed Class~0 objects listed in Table~1 
have L$_{smm}/$L$_{bol} >> 0.5$\% and are likely to be at the beginning 
of the main accretion phase with $M_{env} >> M_\star $ (see Fig.~7b). 

%\placetable{tab1}

%\smallskip
\vskip 0.1truecm
{\it 3. Evolutionary diagrams for embedded YSOs.}\ \
Combining infrared and submillimeter data, it is therefore possible 
to define a complete, empirical evolutionary sequence 
(Class~0~$\rightarrow$~Class~I~$\rightarrow$~Class~II~$\rightarrow$~Class~III)
for low-mass YSOs, which likely correspond to
conceptually different stages of evolution: (early) main accretion phase, 
late accretion phase, PMS stars with protoplanetary disks, PMS stars 
with debris disks (see AM94). 
This sequence is quasi-continuous
and may be parameterized by the ``bolometric temperature'', $T_{bol}$, 
defined by Myers \& Ladd (1993) as the temperature of a blackbody 
having the same mean frequency as the observed YSO SED. 
%In broad terms, 
%$T_{bol}$ measures the redness of a YSO system, including star, 
%disk, and envelope, and can thus be used to track the gradual dissipation of 
%circumstellar material from deeply embedded protostars (low $T_{bol}$)
%to fully revealed PMS stars (high $T_{bol}$). 
%Along these lines,  
Myers \& Ladd proposed to use the $L_{bol}$--$T_{bol}$ diagram 
for embedded YSOs as a direct analog to the H--R diagram for 
optically visible stars. As shown by 
Chen et al. (1995, 1997), YSOs with known classes have distinct  
ranges of $T_{bol}$: $< 70$~K for Class~0, 
70--650~K for Class~I, 650--2880~K for Class~II, and $> 2880$~K for 
Class~III (e.g. Fig.~7a). 
The evolution of $T_{bol}$ and $L_{bol}$ from the Class~0 stage to 
the zero-age main sequence has been modelled in the context of 
various envelope-dissipation scenarios by Myers et al. (1998). 

A perhaps more direct approach to tracking the circumstellar evolution
of YSOs 
%the dissipation of YSO envelopes through accretion and outflow 
is to use the circumstellar mass $M_{c\star}$  
derived from (sub)millimeter continuum measurements of 
optically thin dust emission. Such 
measurements show that $M_{c\star}$ ($= M_{env} + M_{disk}$) is generally 
dominated by $M_{env}$ in Class~0/Class~I sources 
(e.g. Terebey et al. 1993) and decreases by a factor $\sim$~5--10 
on average from one YSO class to the next (AM94). 
In the spirit of the L$_{smm}/$L$_{bol}$ 
evolutionary indicator of AWB93, Saraceno et al. (1996a) proposed the 
$L_{smm}$--$L_{bol}$ (or equivalently $M_{env}$--$L_{bol}$) 
diagram as an alternative evolutionary diagram for self-embedded YSOs. 
%This diagram provides a good way of distinguishing between 
%Class~0 and Class~I objects within large source samples. 
While L$_{smm}$ and $M_{env}$ are well correlated 
with L$_{bol}$ for the majority of embedded YSOs (e.g. Reipurth et al. 1993),
Class~0 objects clearly stand out from Class~I sources in this diagram 
as objects with excess (sub)millimeter emission, i.e., excess 
circumstellar material (see Fig.~7b). 
Moreover, one may compare the locations of observed embedded sources 
in the $M_{env}$--$L_{bol}$ diagram with simple protostellar evolutionary 
tracks (Saraceno et al. 1996a,b). 
%The indicative evolutionary tracks 
%shown in Fig.~7b have been computed assuming that the bolometric luminosities  
%of Class~I and Class~0 protostars derive entirely from accretion, and that
%YSOs form from bounded cloud fragments of finite masses $M_{cl}$ (see II--E). 
%The mass-accretion rate and the envelope mass have also been assumed 
%to decline exponentially with time according to 
%$\maccm = \menvm/\tau = M_{cl}\, exp(-t/\tau)/\tau $, where $\tau = 10^5$~yr. 
Qualitatively at least, scenarios in which 
the mass-accretion rate 
%and/or the rate of envelope dissipation 
decreases with time for a given protostar 
(Bontemps et al. 1996a, Myers et al. 1998 -- see also III--D below) 
and increases with the mass 
%$M_{cl}$ 
of the initial pre-collapse fragment 
(e.g. Myers \& Fuller 1993, Reipurth et al. 1993, Saraceno et al. 1996b) 
yield tracks in better agreement with observations than the constant-rate 
scenario discussed by Saraceno et al. (1996a). 
%(compare our Fig.~7b with Fig.2 of Saraceno et al. 1996a -- 
%see also Myers et al. 1998).
In particular, the peak accretion luminosity  
%reached when half of the initial fragment mass has
%been accreted (hence on the conceptual border line between the Class~0 and 
%Class~I stages -- see Fig.7b), 
is reduced by a factor $\sim$~2--4  
compared to the constant-rate scenario (cf. Bontemps et al. 1996a, 
Myers et al. 1998, and Fig.~7b),
%(e.g. Myers et al. 1998), 
which agrees better 
with observed luminosities (e.g. Kenyon \& Hartmann 1995).

%\placefigure{fig7}

Inclination effects may 
a priori 
affect the positions of individual protostellar objects 
in these evolutionary  
%$L_{bol}$-$L_{smm}$ and $L_{bol}$-$T_{bol}$ 
diagrams.
%Indeed, YSOs surrounded by highly anisotropic distributions of 
%circumstellar matter, such as dense disks, can have their intrinsic 
%bolometric luminosity/temperature severely underestimated 
%when they are viewed close to edge-on (e.g. Kenyon, Calvet, \& Hartmann 1993a, 
%Yorke, Bodenheimer, \& Laughlin 1995, Sonnhalter, Preibisch, and 
%Yorke 1995, Men'shchikov \& Henning 1997). 
%Accordingly, 
In particular, it has been claimed that some Class~I sources may potentially 
look like Class~0 objects when observed at high inclination angles 
to the line of sight (e.g. Yorke, Bodenheimer, \& Laughlin 1995, 
Sonnhalter, Preibisch, \& Yorke 1995, Men'shchikov \& Henning 1997).
%
%\noindent
However, the fact that Class~0 objects are associated with 
an order of magnitude 
more powerful outflows than Class~I sources (see III--D below) confirm that 
these two types of YSOs differ qualitatively from each other.  
%independently of viewing angle. 
Furthermore, some Class~0 sources are 
known to have small inclination angles (e.g. Cabrit \& Bertout 1992, 
Greaves et al. 1997, Wolf-Chase et al. 1998). 
We also stress that existing (sub)millimeter maps of dust continuum 
and C$^{18}$O emission 
provide direct evidence that both Class~I and Class~0 objects 
are {\it self-embedded} in substantial amounts of circumstellar 
material distributed in spatially resolved, spheroidal envelopes  
(e.g. AM94, Chen et al. 1997, Ladd et al. 1998, Dent et al. 1998 -- 
see also III--B below). 
This material has the ability 
to absorb the optical and near-IR emission from the underlying star/disk 
system and to reradiate it quasi-isotropically at longer far-IR and (sub)mm 
wavelengths. 
In such a self-embedded configuration, viewing-angle effects  
are minimized, as confirmed by radiative transfer 
calculations (e.g. Efstathiou \& Rowan-Robinson 1991, Yorke et al. 1995).
Physically, this is because the bulk of the luminosity emerges at 
long wavelengths where most of the emission is effectively optically thin. 
However, the short-wavelength emission from the inner star/disk remains 
very dependent on viewing angle, implying that $T_{bol}$ estimates 
should be somewhat more sensitive 
to orientation effects than L$_{bol}$ and L$_{smm}/$L$_{bol}$.

\vskip 0.1truecm
{\it 4. Protostar surveys and lifetime estimates.}\ \  
Based on the key attributes of Class~0 protostars (see 2. above), 
various strategies can be used to search for and 
discover new candidates: (sub)millimeter 
continuum mapping (e.g. Mezger et al. 1992, Casali et al. 1993, 
Sandell \& Weintraub 1994, 
Reipurth et al. 1996, Launhardt et al. 1996, Chini et al. 1997ab, 
Motte et al. 1998, Andr\'e et al. 1999), 
HIRES processing of the IRAS data (Hurt \& Barsony 1996, 
O'Linger et al. 1999), 
deep radio continuum VLA surveys (e.g. Leous et al. 1991, 
Bontemps et al. 1995, Yun et al. 1996, Moreira et al. 1997, Gibb 1999), 
CO mapping (e.g. Bachiller et al. 1990, Andr\'e et al. 1990, 
Bourke et al. 1997), and large-scale near-IR/optical imaging of 
shocked H$_2$ and [SII] emission (e.g. Hodapp \& Ladd 1995, 
Yu, Bally, \& Devine 1997, Wilking et al. 1997, G\'omez et al. 1998, 
Stanke et al. 1998, Phelps \& Barsony 1999). 

%Table~1 summarizes the main properties of confirmed 
%Class~0 objects which fulfil {\it all} 
%three criteria listed in 2. above. In particular, 
%these objects have all been mapped in the (sub)millimeter continuum.  

%The lifetime of the Class~0 phase is difficult to ascertain with good  
%accuracy since no complete, unbiased census of Class~0 protostars 
%is available yet in any star-forming region.  
%It is, however, clear that 
Class~0 objects appear to be short-lived compared to 
both pre-stellar cores/fragments (see Sect.~II above) 
and Class~I near-IR sources. For example, 
in the $\rho$~Oph main cloud where the IRAM 
30~m mapping study of Motte et al. (1998) 
provides a reasonably complete census of both pre- and proto-stellar 
condensations    
down to $ \simlt 0.1\ M_\odot$, there are only two good 
Class~0 candidates (including the prototypical object VLA~1623 -- AWB93), 
while there are $\sim$~15--30 near-IR Class~I sources 
(e.g. WLY89, AM94, Greene et al. 1994, Barsony \& Ressler 1999, 
Bontemps et al. 1999). 

%\noindent
Under the assumption that $\rho$~Oph 
is representative and forming stars at a roughly constant rate, 
the lifetime of Class~0 objects should thus be approximately an order of 
magnitude shorter than the lifetime of Class~I sources (see 1. above), 
i.e., typically $\sim$~1--3~$ \times 10^4$~yr. The jet-like morphology 
and short dynamical timescales of Class~0 outflows are consistent with this
estimate (e.g. Barsony et al. 1998, see III--D below). 
A lifetime as short as a few $10^4$~yr supports the interpretation of 
Class~0 objects as very young accreting protostars (see, e.g., 
Fletcher \& Stahler 1994 and Barsony 1994). 

Regions like the Serpens, Orion, or Perseus/NGC1333 complexes  
seem to be particularly rich in Class~0 objects, and this is 
probably indicative of fairly high levels of ongoing, likely induced star 
formation activity (e.g. Hurt \& Barsony 1996, Chini et al. 1997b, Yu et al. 1997, Barsony et al. 1998).
%For instance, recent studies of the 
%OMC-3 cloud in Orion~A reveal a filament-like structure, $\simlt 1$~pc 
%long but only $< 0.07$~pc wide, which contains more than half a dozen 
%Class~0 protostars (Chini et al. 1997, Yu et al. 1997). Apparently,  
%a `microburst' of star formation has been triggered in this filament 
%due to external compression by an expanding superbubble 
%powered by the Orion OB association. 
For instance, we estimate the current 
star formation rate in Orion-OMC-3 to be  
$\sim 2 \times 10^{-3}\ M_\odot$~yr$^{-1}$, which is 1--2 orders of 
magnitude larger than the star formation rates characterizing the 
Trapezium, NGC~1333, and IC~348 near-IR clusters when averaged over 
$\simgt 10^6$~yr periods (see Lada, Alves, \& Lada 1996).

%\noindent
%The Serpens and Perseus/NGC1333 complexes also seem to be quite rich in 
%Class~0 objects (e.g. Casali et al. 1993, Hurt \& Barsony 1996, 
%Sandell et al. 1991, 1994, Lefloch et al. 1998, Barsony et al. 1998), 
%which is indicative of 
%fairly high levels of ongoing, likely induced star formation activity. 

%\noindent
%In section IV below, we will suggest that a Class~0 phase of vigorous 
%accretion/ejection may be the natural and inevitable outcome of induced 
%protostellar collapse.

\subsection{B.~~Density Structure of the Protostellar Environment}
%\subsection{B.~~Density Structure of Protostellar Envelopes} 
%
%

{\it 1. Envelope.}\ \ 
In contrast to {\it pre-stellar} cores, the envelopes of  
low-mass Class~0 and Class~I {\it protostars}  
are always found to be strongly centrally-condensed and 
do {\it not} exhibit any inner flattening in their (sub)millimeter 
continuum radial intensity profiles. 
%As a rough observational rule of thumb, 
In practice, this means that, when protostars are 
mapped with the resolution of the largest single-dish telescopes, 
the measured peak flux density is typically a fraction 
$\simgt 20\% $ of the flux integrated over five beam widths. 
%(even after subtraction of a potential unresolved disk component). 
For comparison, the same fraction is $\simlt 10\% $ for pre-stellar cores. 
%in their density profiles. 
%(Using the same notation as in II--C, this means that protostellar envelopes 
%typically have $R_{flat} << 1000$~AU.) 
(Sub)millimeter continuum maps   
indicate that protostellar envelopes 
in regions of isolated star formation such as Taurus 
have radial density gradients  generally consistent with 
$\rho(r) \propto r^{-p}$ with $p \sim $~1.5--2 
over more than $\sim $~10000--15000~AU in radius 
(e.g. Walker, Adams, \& Lada 1990, 
Ladd et al. 1991, Motte 1998, Motte et al. 1999, see also
Mundy et al. this volume). The estimated density gradient 
thus agrees with most  
collapse models which predict a value of $p$ between 1.5 and 2  
during the protostellar accretion phase (e.g. Whitworth \& Summers 1985). 
%A few studies, primarily based on modelling of the SEDs, 
Some studies 
have, however, inferred shallower density gradients ($p \sim $~0.5--1 --
e.g. Barsony \& Chandler 1993, Chandler, Barsony, \& Moore 1998). 
In any case, the densities and masses measured for the envelopes around 
the bona-fide Class~I objects of Taurus 
appear to be consistent within a factor of $\sim 4$   
with the predictions of the ``standard'' inside-out collapse theory 
(e.g. Shu et al. 1993) for $\sim 10^5$~yr-old, isolated protostars 
(Motte 1998, Motte et al. 1999). 
%This simple situation does not hold, however, 

%\noindent
The situation is markedly different in star-forming clusters. 
In $\rho$ Ophiuchi in particular, the circumstellar envelopes of  
Class~I and Class~0 protostars are observed to be very compact: 
they merge with dense cores, other envelopes, 
and/or the diffuse ambient cloud at a {\it finite} 
radius $R_{out} \simlt 5\,000$~AU (Motte et al. 1998). 
This is $\simgt$~3 times smaller than the collapse 
expansion wavefront at a `Class~I age' of $ \sim 2 \times 10^5$~yr in the 
standard theory of isolated protostars, emphasizing the fact 
that each YSO has a finite `sphere of influence' in $\rho$ Oph. 
Similar results were obtained in the case of 
the Perseus Class~0 sources NGC1333-4A, NGC1333-2, 
L1448-C, and L1448-N (e.g. Motte 1998).  
Moreover, the envelopes of these Perseus protostars are  
3 to 10 times denser than the singular isothermal sphere for a sound 
speed $a = 0.2\ \kms $. This suggests that, prior to collapse, the main 
support against gravity was turbulent and/or magnetic in origin rather than 
purely thermal (see also Mardones et al. 1997).
%In agreement with this view, they generally exhibit 
%broad linewidths indicative of nonthermal velocity dispersions 
%(e.g. Mardones et al. 1997).

%\smallskip
\vskip 0.1truecm
{\it 2. Disks and multiplicity.}\ \ 
Many Class~0 protostars are in fact multiple systems, when viewed at 
sub-arcsecond resolution, sharing a common envelope and sometimes 
a circumbinary disk (e.g. Looney et al. 1999 -- see 
also col.~11 of Table~1 and chapter by Mundy, Looney, \& Welch). 
These protobinaries probably formed by dynamical fragmentation 
during (or at the end of) the isothermal collapse phase 
(e.g. Chapman et al. 1992, Bonnell 1994, Boss \& Myhill 1995). 
Interestingly enough, only cores with inner density profiles 
as flat as $\rho \propto r^{-1}$ or flatter 
(like observed pre-stellar cores -- see Sect.~II), 
%only moderately centrally-condensed cloud cores, resembling observed 
%pre-stellar cores (see Sect.~II above), 
can apparently fragment 
during collapse (Myhill \& Kaula 1992, Boss 1995, Burkert et al. 1997).

%\noindent
Despite the difficulty of discriminating between the disk 
and envelope components, existing (sub)millimeter continuum 
interferometric measurements suggest that 
the ``disks'' of Class~0 objects are a factor of 
$\simgt 10$ less massive 
than their surrounding circumstellar envelopes (e.g.  
Chandler et al. 1995, Pudritz et al. 1996, 
Looney et al. 1999, Hogerheijde 1998, Motte 1998, 
Wilner \& Lay this volume).  
%By contrast, the disk is a very significant 
%(sometimes dominant) component in more evolved (Class~I and Class~II) sources
%(e.g. Lay et al. 1994). It is, however, not clear at present 
%whether the disk systematically grows in size and mass with time.
%

\subsection{C.~~Direct Evidence for Infall}
Rather convincing spectroscopic signatures of gravitational infall 
have been reported toward several Class~0 objects, 
confirming their protostellar nature (e.g. Walker et al. 1986, 
Zhou et al. 1993, Gregersen et al. 1997, Mardones et al. 1997, 
see also col. 10 of Table~1).  
Inward motions can be traced by optically thick molecular lines 
which should (locally) exhibit 
asymmetric self-absorbed profiles skewed to the blue 
(see Myers, Evans, \& Ohashi, this volume).  
The interpretation is often complicated by the
simultaneous presence of 
rotation and/or outflow (e.g. Menten et al. 1987, 
Walker et al. 1994, Cabrit et al. 1996).
%

%\noindent 
A comprehensive survey of a sample of 47 embedded YSOs in 
H$_2$CO$(2_{12}-1_{11})$ and CS(2--1) suggests that 
infall is more prominent in Class~0 than in Class~I sources 
(Mardones et al. 1997): In these transitions, infall 
asymmetries are detected toward 40--50~\% of Class~0 objects but 
less than 10~\% of Class~I sources. 
This is qualitatively consistent with a decline of infall/accretion rate 
with evolutionary stage (see III--D below).
However, a more recent survey by Gregersen et al. (1999) 
using HCO$^+$(3--2) finds no difference 
in the fraction of sources with ``blue profiles'' between 
Class~0 and Class~I sources. The HCO$^+$(3--2) line is more optically 
thick than H$_2$CO$(2_{12}-1_{11})$ and CS(2--1), hence a better tracer 
of infall at advanced stages. This result shows 
that some infall is still present at the Class~I stage but 
remains consistent with a decline of the net accretion rate with time.  
The outflow is so broad in Class~I sources that there often appears to be 
little transfer of mass to the inner $\sim 2000$~AU radius region around 
these objects (e.g. Fuller et al. 1995b, Cabrit et al. 1996, 
Brown \& Chandler 1999). 
%for detailed studies of L1551-IRS5 and HL~Tau, respectively). 

%\subsection{C.~~Outflow/Jet Properties} 
%\subsection{D.~~Decline of Mass Loss and Mass Accretion with Time} 
%\subsection{B.~~Decline of Outflow Power with Time}
\subsection{D.~~Decline of Outflow and Inflow with Time}

{\it 1. Evolution from Class~0 to Class~I.}\ \
Most, if not all, Class~0 protostars drive powerful,  
``jet--like'' CO molecular outflows (see, e.g., Bachiller 1996 
%Cabrit, Raga, \& Gueth 1997, 
and Richer et al., this volume, for reviews).  
The mechanical luminosities of these outflows   
are often of the same order as the bolometric luminosities of the 
central sources (e.g. Curiel et al. 1990, AWB93, Barsony et al. 1998).
In contrast, while there is good evidence that 
some outflow activity exists throughout the accretion phase 
(e.g. Terebey et al. 1989, Parker et al. 1991, Bontemps et al. 1996a),  
the CO outflows from Class~I sources tend to be much less powerful 
and less collimated than those from Class~0 objects.

In an effort to quantify the evolution of mass loss during the 
protostellar phase, Bontemps et al. (1996a -- hereafter BATC) obtained and
analyzed a homogeneous set of CO(2--1) outflow data around a large sample of 
low-luminosity ($L_{bol} < 50\ L_\odot $), nearby ($d < 450$~pc) 
self-embedded YSOs. 
Their results show that Class~0 objects lie an order of magnitude 
above the well-known (e.g. Cabrit \& Bertout 1992) 
correlation between outflow momentum flux ($F_{\rm CO}$) and 
bolometric luminosity ($L_{\rm bol}$) holding for Class~I sources  
(see \fco--\lbol diagram shown in Fig.~5 of BATC). 
Furthermore, BATC found that 
%Figure~\ref{norm_fco_menv} shows that  
\fco was well correlated with $M_{env}$ in their {\it entire}  
sample (including both Class~I and Class~0 sources). The same correlation 
was noted independently on other source samples by 
Moriarty-Schieven et al. (1994), Hogerheijde et al. (1998), 
and Henning \& Launhardt (1998). 
As argued by BATC, this new correlation is independent of the   
\fco--\lbol correlation and most likely results from a progressive  
decrease of outflow power with time during the accretion phase. 
This is illustrated in the normalized 
$\fcom\,\rm{c}/\lbolm$ versus $\menvm/\lbolm^{0.6}$ diagram of 
%Figure~\ref{norm_fco_menv}, 
Fig.~8, which should be essentially free of any luminosity effect. 

Since magneto-centrifugal accretion/ejection models of bipolar outflows 
(e.g. Shu et al. 1994, Ferreira \& Pelletier 1995, 
Fiege \& Henriksen 1996, Ouyed \& Pudritz 1997) 
predict a direct proportionality between accretion and ejection,   
%that the mass-loss rate is a definite fraction of the mass-accretion rate, 
BATC proposed that the decline of outflow power 
with evolutionary stage seen in Fig.~8   
reflects a corresponding decrease in the mass-accretion/infall rate.  
The results of BATC indicate that \mjet declines from $\sim 10^{-6}\,\myr$ 
for the youngest Class~0 protostars to $\sim 2 \times 10^{-8}\,\myr$ 
for the most evolved Class~I sources, suggesting a decrease in  
\macc from $\sim 10^{-5}\,\myr$ to $\sim 2 \times 10^{-7}\,\myr$  
if realistic jet model parameters are adopted 
($ \mjetm/\maccm \sim $~0.1--0.3, $ \vjetm \sim 100\, \kms $).
These indirect estimates of \macc for Class~0 and Class~I protostars 
%are admittedly not very accurate and 
should only be taken as indicative of the true evolutionary trend.
Nevertheless, it is interesting to note that they agree well   
with independent estimates of the rates of envelope dissipation 
based on circumstellar mass versus age arguments (Ward-Thompson 1996, 
Ladd, Fuller, \& Deane 1998).
%simple envelope-mass/lifetime arguments 
%and an analysis of the envelope mass distributions observed by AM94 
%(Ward-Thompson 1996) and with the rates of envelope dissipation 
%inferred by Ladd, Fuller, \& Deane (1998). 
%They also are in rough agreement with the rates of envelope dissipation 
%inferred by Ladd, Fuller, \& Deane (1998) for their sample of embedded YSOs  
%in Taurus. (According to Ladd et al., however, the dissipation 
%of the envelope occurs primarily through outflow at early times 
%and through accretion at late times.)

%\noindent
As illustrated by the evolutionary tracks of Fig.~7b, a decline of \macc 
with time does not imply a higher accretion luminosity for 
Class~0 objects   
compared to Class~I sources because the central stellar mass is
smaller at the Class~0 stage and the stellar radius is likely to be larger
%, and both the stellar radius and the amount of 
%accretion energy dissipated in the wind are likely to be larger 
(see Henriksen, Andr\'e, \& Bontemps 1997 -- hereafter HAB97). 

%\placefigure{fig8}
 
\vskip 0.1truecm
{\it 2. Link with the collapse initial conditions.}\ \
%We may go further and link 
The apparent decay of \macc from the Class~0 to the Class~I stage 
may be linked with the density structure 
observed for pre-stellar cores/condensations (Sect.~II). 
%In the standard theory of isolated low-mass star 
%formation (e.g. Shu et al. 1993), the initial conditions 
%correspond to singular isothermal spheres
%(SISs with $\rho = (a^2/2\pi G )r^{-2}$), and    
%this leads to a strictly constant accretion rate, $\maccm \sim a^3/G$,
%where $a$ is the isothermal sound speed (Shu 1977). 
%However, 
(Magneto)hydrodynamic collapse models 
predict a time-dependent accretion history 
%when the collapse initial conditions are either not singular or not scale-free
when the radial density profile at the onset of fast protostellar collapse 
differs from $\rho \propto r^{-2}$ 
(e.g. Foster \& Chevalier 1993 -- FC93, Tomisaka 1996, 
McLaughlin \& Pudritz 1997, HAB97, 
Basu 1997, Safier, McKee, \& Stahler 1997, Li 1998, Ciolek \& K\"onigl 1998). 
In particular, when starting from Bonnor-Ebert-like initial 
conditions resembling observed 
pre-stellar cores, these studies find that supersonic infall 
velocities develop prior to the formation of the central  
hydrostatic protostellar object at $t=0$ (see, e.g., FC93).
Observationally, this early collapse phase  
%which does not exist in the 
%standard theory of isolated low-mass star formation (Shu et al. 1987, 1993), 
should correspond to `isothermal protostars' (see II--E for possible 
examples). During the protostellar accretion phase ($t > 0$), 
because of the significant infall velocities achieved at $t < 0$, 
\macc is initially higher than the standard $\sim a^3/G$ value obtained 
for the inside-out collapse of a static singular isothermal sphere 
(Shu 1977, see also Sect.~I). 
%where $a$ is the isothermal sound speed (Shu 1977).
The accretion rate then converges toward the standard value  
of the Shu solution, and declines again below $a^3/G$ at late times if the 
reservoir of mass is finite. 
By comparison with the rough estimates of \macc given above, 
it is tempting to identify the short period of energetic accretion 
($\maccm \sim 10 \times a^3/G $) 
predicted by the models just after point-mass formation   
with the observationally-defined Class~0 stage 
(HAB97, see Fig.~8). 
In this view, the more evolved Class~I objects would correspond to 
the longer period of moderate accretion/ejection when the accretion rate 
approaches the standard value 
($\maccm \simlt a^3/G \sim 2 \times 10^{-6}\, \myr$ for a cloud temperature 
of $\sim 10$~K).

\noindent
%Because the inflow becomes supersonic early on, one may obtain a qualitatively 
%correct description of the collapse and subsequent accretion phase 
%through simple, pressure-free analytical calculations (HAB97).  
Using simple pressure-free analytical calculations
%their simple model 
(justified since the inflow becomes supersonic early on), 
HAB97 could indeed find a good overall fit to 
the empirical accretion history inferred by BATC 
on the basis of CO outflow observations (see Fig.~8). 
%when a relatively large fraction ($\sim 30\ \% $) of the initial mass 
%lies in the uniform-density central region of the model  
%pre-collapse condensation.

In the absence of magnetic fields, FC93 have 
shown that the timescale for convergence to the standard accretion 
rate of Shu (1977) depends on the radius of the flat inner core 
($R_{flat}$) relative to the outer radius of the initial pre-collapse 
condensation ($R_{out}$). They found that a phase of constant $\sim a^3/G$ 
accretion rate is achieved only when $R_{out}/R_{flat} \simgt 20$, typically after $\sim 10$ free-fall times of the flat inner region, for a period 
lasting $\sim 15$ free-fall times when $R_{out}/R_{flat} = 20$ and 
progressively longer as $R_{out}/R_{flat}$ increases. 
Since observations suggest $R_{out}/R_{flat} \simlt 3$ in Ophiuchus 
and $R_{out}/R_{flat} \simgt 15$ in Taurus (see II--C, III--B, and 
Motte et al. 1998), one may expect a marked time-dependence of \macc
in Ophiuchus but a reasonable agreement with the 
constant accretion rate of the self-similar theory of 
Shu et al. (1987, 1993) in Taurus. Indeed, 
HAB97 note that there is a much better continuity 
between Class~0 and Class~I protostars 
%, and an apparently more gentle time-dependence of \macc, 
in Taurus than in Ophiuchus (see also Andr\'e 1997).
 
Finally, we stress that 
the absolute values of \macc in the Class~0 and Class~I stages 
are presently quite uncertain.
An alternative interpretation of the evolution seen in Fig.~8 is 
that Class~0 protostars accrete at a rate roughly consistent 
with $\sim a^3/G$, while most Class~I sources are in a terminal accretion 
phase with $\maccm \simlt 0.1 \times a^3/G$, resulting from  
the finite effective reservoir of mass available to each object in clusters
(e.g. Motte et al. 1998 and III--B) 
and/or from the effects of outflows dispersing the envelope 
(e.g. Myers et al. 1998, Ladd et al. 1998). Distinguishing between this 
possibility and that advocated above 
will require direct measurements of the mass accretion rates.

\mainsection{{I}{V}.~~CONCLUSIONS AND IMPLICATIONS}
\backup
The observational studies discussed in Sect.~II demonstrate that
pre-stellar cores/fragments are characterized by flat inner radial 
density gradients. This, in turn, suggests the initial conditions for fast 
protostellar collapse are 
%often 
non-singular, i.e., the density 
profile at the onset of collapse is not infinitely centrally condensed.  
(Sub)millimeter observations also set strong constraints on protostellar 
evolution (Sect.~III). The fact that young (Class~0) protostars drive more 
powerful outflows 
%and are more often associated with spectroscopic infall signatures 
than evolved (Class~I) protostars suggests that the 
mass accretion rate \macc decreases by typically a factor of $\sim$~5--10 
%one order of magnitude 
from the Class~0 to the Class~I stage (III--D). 
Such a decline in \macc  
during the protostellar accretion phase may be the direct result of 
a flattened initial density profile (see III--D).
%As we discuss in IV--B below, 
%We now bring these two sets of observational constraints together and argue  
Based on these observational constraints, we suggest that most  
protostars form in a dynamical rather than quasi-static fashion. 
  
%
%\subsection{B.~~Origin of the IMF}
 
The results summarized in this chapter also have 
broader implications concerning, e.g., the origin of the IMF.
As pointed out in Sect.~II--E, the pre-stellar condensations observed in 
regions of multiple star formation such as $\rho$~Ophiuchi  
are finite-size structures, typically a few 1000~AU in radius, 
which are clearly not scale-free. 
This favors a picture of star formation in clusters 
in which individual protostellar 
collapse is initiated in compact dense clumps resulting from fragmentation 
and resembling more finite-size Bonnor-Ebert cloudlets than singular 
isothermal spheres.  
Such condensations may correspond to dense, low-ionization pockets 
decoupling themselves from the parent molecular cloud as a result of 
ambipolar diffusion and/or the dissipation of turbulence 
(e.g. Mouschovias 1991, Myers 1998, Nakano 1998). They would thus be free 
to undergo Jeans-like gravitational instabilities and collapse under the 
influence of external disturbances, of which there are many types 
in regions of multiple star formation (e.g. Pringle 1989, 
Whitworth et al. 1996, 
Motte et al. 1998, Barsony et al. 1998). By contrast, 
the low-density regions of the ambient cloud, being 
more ionized, would remain supported against collapse by static and/or 
turbulent magnetic fields. The typical separation between individual 
condensations should be of order the Jeans length in the parent 
cloud/core, in rough agreement with observations (e.g. Motte et al. 1998). 
 
In this observationally-driven scenario of fragmentation and collapse, 
stars are built from bounded fragments 
which represent finite reservoirs of mass.
%Furthermore, most of the fragment initial masses, i.e., 
%the fragment masses at the onset of collapse, end up in stars. 
%Furthermore, 
The star formation efficiency within these 
fragments is high: 
%most of the fragment initial masses end up in stars.
most of their `initial' masses at the onset of collapse end up in stars.
If this is true, it implies that 
%{\it the IMF of embedded clusters 
%is primarily determined at the pre-stellar stage of star formation} and that
the physical mechanisms  
responsible for the formation of pre-stellar cores/condensations  
in molecular clouds, such as turbulent fragmentation 
%(e.g. Padoan et al. 1997),
(e.g. Padoan, Nordlund, \& Jones 1997), 
play a key role in determining the IMF of embedded clusters.
%fixing the stellar mass scale. 
Such a picture, which we favor for regions of multiple star formation,  
%such as $\rho$~Ophiuchi, 
is consistent with some theoretical 
scenarios of protocluster formation (e.g. Larson 1985, 
%Klessen et al. 1998).
Klessen, Burkert, \& Bate 1998). 
It need not be universal, however, and 
is in fact unlikely to apply to regions of isolated star formation like 
the Taurus cloud. 
In these regions, protostars may accrete from larger
reservoirs of mass, and feed-back processes such as stellar winds 
may be more important in limiting accretion and defining stellar masses 
(e.g. Shu et al. 1987, Silk 1995, Adams \& Fatuzzo 1996, 
Velusamy \& Langer 1998). 

%
%\subsection{B.~~Future Prospects} 

With the advent of major new 
%ground-based and space-borne 
facilities 
at far-IR and submillimeter wavelengths, 
the next decade promises to be at least as rich  
in observational discoveries 
%on the earliest stages of star formation 
as the past ten years. 
%In particular,  
By combining the capabilities of space telescopes such as FIRST 
with those of large ground-based arrays such as the LSA/MMA, 
it will be possible to study the detailed physics of complete samples of 
young protostars and pre-collapse fragments, in a variety of star-forming 
clouds, and down to the brown-dwarf mass regime.
%and for 
%the whole spectrum of stellar masses.  
This should tremendously improve our global understanding of the initial 
stages of star formation in the Galaxy.

{\it Acknowledgements.} We thank C. Correia for providing unpublished  
data shown in Fig.~2, G. Ciolek for Fig.~3b, P. Myers for Fig.~7a, 
and N. Grosso for assistance in preparing some of the figures. 
We are also grateful to F. Motte and the referee, N. Evans, for 
useful comments and suggestions.

%\medskip

\vfill\eject
\null

\vskip .5in
\centerline{\bf REFERENCES}
\vskip .25in

\ref{Abergel, A., Bernard, J.P., Boulanger, F. et al. 1996. ISOCAM mapping 
  of the $\rho$ Ophiuchi main cloud. {\refit Astron.\ Astrophys.\/} 
  315:L329--L332}

\ref{Abergel, A., Bernard, J.P., Boulanger, F. et al. 1998. The dense core 
  Oph D seen in extinction by ISOCAM. 
  In {\refit Star Formation with ISO\/}, eds.\ J.L. Yun 
 \& R. Liseau, {\refit A.S.P. Conf. Series\/}, 132:220--229}

\ref{Adams F.C., and Fatuzzo, M. 1996. A theory of the initial mass function
for star formation in molecular clouds.
{\refit Astrophys.\ J.\/}, 464:256--271 }

\ref{Adams F.C., Lada C.J., and Shu, F.H. 1987. Spectral evolution of young stellar
objects. {\refit Astrophys.\ J.\/} 312:788--806 }

\ref{Akeson, R.L., and Carlstrom, J.E. 1997. Magnetic field structure 
  in protostellar envelopes. {\refit Astrophys.\ J.\/} 491:254--266 }

\ref{Andr\'e, P. 1997. The evolution of flows and protostars. 
    In {\refit Herbig-Haro Flows and the Birth of Low Mass Stars. 
    Proc. IAU Symp. 182\/}, 
    eds.\ B. Reipurth and C. Bertout (Dordrecht: Kluwer), pp.\ 483--494. }

\ref{Andr\'e, P., and Montmerle, T. 1994. From T Tauri stars to protostars: 
Circumstellar material and young stellar objects in the $\rho$ Ophiuchi cloud. 
{\refit Astrophys.\ J.\/} 420:837--862 (AM94)}

\ref{Andr\'e, P., Motte, F., and Bacmann, A. 1999. Discovery of an extremely 
  young accreting protostar in Taurus. 
  {\refit Astrophys.\ J.\ Lett.\/}, in press }

\ref{Andr\'e P., Ward-Thompson D., and Barsony, M. 1993. Submillimeter 
continuum observations of $\rho$~Ophiuchi~A: The candidate protostar VLA~1623
and prestellar clumps. {\refit Astrophys.\ J.\/} 406:122--141 (AWB93)}

\ref{Andr\'e P., Ward-Thompson D., and Motte, F. 1996. Probing the initial conditions 
of star formation: the structure of the prestellar core L1689B. 
{\refit Astron.\ Astrophys.\/} 314:625--635} 

\ref{Andr\'e, P., Mart\'\i n-Pintado, J., Despois, D., and  
     Montmerle T. 1990. Discovery of a remarkable bipolar flow and 
     exciting source in the $\rho$ Ophiuchi cloud core. 
     {\refit Astron.\ Astrophys.\/} 236:180--192}

\ref{Anglada, G., Estalella, R., Rodr\'\i guez, L.F., Torrelles, J.M., 
L\'opez, R., \& Cant\'o, J. 1991. A double radio source at the center of the outflow 
in L723. {\refit Astrophys.\ J.\/} 376:615--617}

\ref{Anglada, G., Rodr\'\i guez, L.F., Cant\'o, J., Estalella, R., \& 
  Torrelles, J.M. 1992. Radio continuum from the powering sources of the 
  RNO 43, HARO 4-255 FIR, B335, and PV Cephei outflows and from the 
  Herbig-Haro object 32A. {\refit Astrophys.\ J.\/} 395:494--500}

\ref{Avery, L.W., Hayashi, S.S., and White, G.J. 1990. 
  The unusual morphology of the high-velocity gas in L723 - One outflow 
  or two? {\refit Astrophys.\ J.\/} 357:524--530}

\ref{Bachiller, R. 1996. Bipolar molecular outflows from young stars 
   and protostars.  
  {\refit Ann.\ Rev.\ Astron.\ Astrophys.\/} 34:111--154 }

\ref{Bachiller, R., Andr\'e, P., and Cabrit, S. 1991a. Detection of the 
   exciting 
  source of the spectacular molecular outflow L1448 at $\lambda\lambda$~1-3 mm.
  {\refit Astron.\ Astrophys.\/} 241:L43--L46}

\ref{Bachiller, R., Mart\'\i n-Pintado, J., Tafalla, M., Cernicharo, J., 
   Lazareff, B. 1990. High-velocity molecular bullets in a fast bipolar outflow 
   near L1448/IRS3. {\refit Astron.\ Astrophys.\/} 231:174--186}

\ref{Bachiller, R., Guilloteau, S., Dutrey, A., Planesas, P., 
   Mart\'\i n-Pintado, J. 1995. The jet-driven molecular outflow in L 1448. 
   CO and continuum synthesis images. 
   {\refit Astron.\ Astrophys.\/} 299:857--868}

\ref{Bachiller, R., Guilloteau, S., Gueth, F., Tafalla, M., Dutrey, A., 
   Codella, C., and Catsets, A. 1998. A molecular jet from SVS 13B 
   near HH~7-11. {\refit Astron.\ Astrophys.\/} 339:L49--L52}

\ref{Bachiller, R., Mart\'\i n-Pintado, J., \& Planesas, P. 1991b. 
    High-velocity molecular jets and bullets from IRAS 03282$+$3035. 
   {\refit Astron.\ Astrophys.\/} 251:639--648}
 
\ref{Bachiller, R., Terebey, S., Jarrett, T., Mart\'\i n-Pintado, J., 
  Beichman, C.A., \& van Buren, D. 1994. Shocked molecular gas around 
  the extremely 
  young source IRAS 03282$+$3035.{\refit Astrophys.\ J.\/} 437:296--304}

\ref{Bacmann, A., Andr\'e, P., Abergel, A. et al. 1997. An ISOCAM 
  absorption study of dense cloud cores. 
  In {\refit Star Formation with ISO\/}, eds.\ J.L. Yun 
 \& R. Liseau, {\refit A.S.P. Conf. Series\/}, 132:307--313} 

\ref{Bally, J., Lada, E.A., \& Lane, A.P. 1993. The L1448 Molecular Jet. 
  {\refit Astrophys.\ J.\/} 418:322--327}

\ref{Balsara D., Ward-Thompson D., Pouquet A., Crutcher R. M. 1998. 
 An MHD model 
of the inter-stellar medium and a new method of accretion onto dense star-forming  cores. In {\refit Interstellar Turbulence\/}, eds.\ J. Franco and A. Carraminana A., 
 (Cambridge: University Press), pp.\ 49-- }

\ref{Barsony, M. 1994. Class 0 protostars. In {\refit Clouds, Cores, 
 and Low-mass Stars\/}, eds.\ D. P. Clemens and R. Barvainis, 
 {\refit A.S.P. Conf. Series\/}, 65:197--206 }

\ref{Barsony, M., and Chandler, C. J. 1992. 
  On the origin of submillimeter emission from young stars in Taurus-Auriga. 
  {\refit Astrophys.\ J.\/} 384:L53--L57 }

\ref{Barsony, M., and Chandler, C. J. 1993. 
  The circumstellar density distribution 
  of L1551NE. {\refit Astrophys.\ J.\/} 406:L71--L74 }

\ref{Barsony, M., Ressler, M. 1999. The initial luminosity function for 
  L1688: New mid-IR imaging photometry. {\refit Astrophys.\ J.\/},
  in preparation }

\ref{Barsony, M., Ward-Thompson, D., Andr\'e, P., and O'Linger, J. 1998. 
  Protostars in Perseus: Outflow induced fragmentation. 
 {\refit Astrophys.\ J.\/}  509:733--748 }

\ref{Basu, S. 1997. A Semianalytic Model for Supercritical Core Collapse: Self-similar 
  Evolution and the Approach to Protostar Formation. 
  {\refit Astrophys.\ J.\/} 485:240--253}

\ref{Basu, S., and Mouschovias, T.Ch. 1995. Magnetic Braking, Ambipolar Diffusion, 
  and the Formation of Cloud Cores and Protostars. III. Effect of the Initial 
  Mass-to-Flux Ratio. {\refit Astrophys.\ J.\/} 453:271--283}

\ref{Bate, M.R. 1998. Collapse of a Molecular Cloud Core to Stellar Densities: 
  The First Three-dimensional Calculations. {\refit Astrophys.\ J.\ Lett.\/} 
   508:L95--L98 }

\ref{Beckwith, S.V.W., Sargent A.I, Chini, R.S., Guesten, R. 1990. A survey for 
  circumstellar disks around young stellar objects. {\refit Astron.\ J.\/} 99:924--945}

\ref{Beichman, C.A., Myers, P.C., Emerson, J.P. et al. 1986. Candidate 
  solar-type protostars in nearby molecular cloud cores. 
  {\refit Astrophys.\ J.\/} 307:337--349}

\ref{Bence, S.J., Richer, J.S., and Padman, R. 1996. RNO 43: a jet-driven 
  super-outflow. {\refit Mon.\ Not.\ Roy.\ Astron.\ Soc.\/} 279:866--883}

\ref{Benson, P. J., and Myers, P. C. 1989. A survey for dense cores in dark clouds.
  {\refit Astrophys.\ J.\ Sup.\/} 71:89--108}

\ref{Beichman, C.A., Myers, P.C., Emerson, J.P. et al. 1986. Candidate solar-type 
  protostars in nearby molecular cloud cores. {\refit Astrophys.\ J.\/} 307:337--349}

\ref{Blake, G.A., Sandell, G., van Dishoek, E.W., Groesbeck, T.D., 
     Mundy, L.G., and Aspin, C. 1995. 
     A molecular line study of NGC 1333/IRAS 4.  
     {\refit Astrophys.\ J.\/} 441:689--701}

\ref{Blitz, L. 1993. Giant molecular clouds. In {\refit Protostars 
  \& Planets {I}{I}{I}\/}, eds.\ E. H. Levy and J. I. Lunine (Tucson: Univ.\ 
  of Arizona  Press), pp.\ 125--161.}

\ref{Blottiau, P., Chi\`eze, J.P., and Bouquet, S. 1988. An asymptotic 
   self-similar solution for the gravitational collapse 
   {\refit Astron.\ Astrophys.\/} 207:24--36 }

\ref{Brown, D. W., and Chandler, C.J. 1999. Circumstellar kinematics and 
   the measurement of stellar mass for the protostars TMC1A and TMC1A. 
   {\refit Mon.\ Not.\ Roy.\ Astron.\ Soc.\/} April/May}

\ref{Bonnell, I. A. 1994. Fragmentation and the formation of binary and 
  multiple systems. In {\refit Clouds, Cores, and Low-mass Stars\/}, 
  eds.\ D. P. Clemens and R. Barvainis, {\refit A.S.P. Conf. Series\/}, 
  65:115--124 }

\ref{Bonnell, I. A., Bate, M. R., Clarke, C. J. and Pringle, J.E. 1997.
  Accretion and the stellar mass spectrum in small clusters. 
  {\refit Mon.\ Not.\ Roy.\ Astron.\ Soc.\/} 285:201--208}

\ref{Bontemps, S. 1996. Evolution de l'\'ejection de mati\`ere des proto-\'etoiles.
  {\refit Ph.\ D.\ Dissertation,\ University\ of\ Paris\ XI\/} }

\ref{Bontemps, S., Andr\'e, P., and Ward-Thompson, D. 1995. Deep VLA search for 
  the youngest protostars: A Class 0 source in the HH24-26 region. 
  {\refit Astron.\ Astrophys.\/} 297:98--102 }

\ref{Bontemps, S., Andr\'e, P., Terebey, S., and Cabrit, S. 1996a. Evolution of 
  outflow activity around low-mass embedded young stellar objects. 
  {\refit Astron.\ Astrophys.\/} 311:858--872 (BATC) }

\ref{Bontemps, S., Nordh, L., Olofsson, G. et al. 1999. 
  An ISOCAM Deep Census of Low-Mass Stars in $\rho$~Ophiuchi -- Mass 
  Function in a Young Embedded Cluster. 
  In {\refit The Universe as seen by ISO\/, eds.\ P. Cox and M. Kessler 
  (Noordwijk: ESA-SP), in press }

\ref{Bontemps, S., Ward-Thompson, D., and Andr\'e, P. 1996b. 
  Discovery of a jet emanating from the protostar HH 24 MMS. 
  {\refit Astron.\ Astrophys.\/} 314:477--483 }

%\ref{Boss, A.P. 1988. Binary stars: Formation by fragmentation
%{\refit Comments.\ Astrophys.\/} 12:169--190}

\ref{Boss, A.P. 1995. Gravitational Collapse and Binary Protostars.
{\refit Rev.\ Mex.\ A.\ A.\ (ser. de conf.)\/} 1:165--177}

\ref{Boss, A.P., \& Myhill, E.A. 1995. Collapse and Fragmentation of 
  Molecular Cloud Cores. III Initial Differential Rotation.
  {\refit Astrophys.\ J.\/} 451:218--224}

\ref{Boss, A.P., \& Yorke, H.W. 1995. Spectral energy of first protostellar cores: 
  Detecting 'class -I' protostars with ISO and SIRTF. {\refit Astrophys.\ J.\/} 
  439:L55--L58}

\ref{Bourke, T.L., Hyland, A.R., \& Robinson, G. 1995a. Studies of star formation 
  in isolated small dark clouds - I. A catalogue of southern Bok globules: optical 
  and IRAS properties. {\refit Mon.\ Not.\ Roy.\ Astron.\ Soc.\/} 276:1052-1066}

\ref{Bourke, T.L., Hyland, A.R., \& Robinson, G. 1995b. Studies of star formation 
  in isolated small dark clouds - II. A southern ammonia survey. 
  {\refit Mon.\ Not.\ Roy.\ Astron.\ Soc.\/} 276:1067-1084}

\ref{Bourke, T. L., Garay, G., Lehtinen, K. K., et al. 1997. Discovery of 
  a Highly Collimated Molecular Outflow in the Southern Bok Globule BHR 71.
  {\refit Astrophys.\ J.\/} 476:781--800 }

\ref{Brown, D.W., and Chandler, C.J. 1999. Circumstellar kinematics and 
  the measurement of stellar mass for the protostars TMC1 and TMC1A. 
   {\refit Mon.\ Not.\ Roy.\ Astron.\ Soc.\/} April/May}

\ref{Burkert, A., Bate, M.R., and Bodenheimer, P. 1997. Protostellar  
  fragmentation in a power-law density distribution. 
  {\refit Mon.\ Not.\ Roy.\ Astron.\ Soc.\/} 289:497-504}

\ref{Butner, H.M., Lada, E.A., and Loren, R.B. 1995. Physical 
   Properties of Dense Cores: DCO + Observations.
  {\refit Astrophys.\ J.\/} 448:207--225 }

\ref{Cabrit, S., \& Andr\'e, P. 1991. An observational connection between 
  circumstellar disk mass and molecular outflows. 
  {\refit Astrophys.\ J.\ Lett.\/} 379:L25--L28}

\ref{Cabrit S., \& Bertout C. 1992. CO line formation in bipolar flows. 
  III - The energetics of molecular flows and ionized winds. 
  {\refit Astron.\ Astrophys.\/} 261:274--284}

\ref{Cabrit S., Goldsmith, P.F., and Snell, R.L. 1988. Identification of 
  RNO 43 and B335 as two highly collimated bipolar flows oriented nearly 
  in the plane of the sky. 
  {\refit Astrophys.\ J.\/} 334:196--208}

\ref{Cabrit, S., Guilloteau, S., Andr\'e, P., Bertout, C., Montmerle, T., 
   \& Schuster, K. 1996. Plateau de Bure observations of HL Tauri: 
   outflow motions in a remnant circumstellar envelope. 
   {\refit Astron.\ Astrophys.\/} 305:527--540}

\ref{Casali, M. M., Eiroa, C., and Duncan, W. D. 1993. A second phase of star 
   formation in the Serpens core. {\refit Astron.\ Astrophys.\/} 275:195--200 }

\ref{Ceccarelli, C., Caux, E., White, G. J. et al. 1998. The far infrared line 
   spectrum of the protostar IRAS 16293-2422. {\refit Astron.\ Astrophys.\/} 
   331:372--382 }

\ref{Cernicharo, J., Lefloch, B., Cox, P. et al. 1998. Induced Massive 
   Star Formation in the Trifid Nebula? {\refit Science} 
   282:462--465 }

\ref{Chandler, C.J., Barsony, M., and Moore, T.J.T. 1998. The circumstellar 
   envelopes around three protostars in Taurus. 
   {\refit Mon.\ Not.\ Roy.\ Astron.\ Soc.\/} 299:789--798}

\ref{Chandler, C.J., Gear, W.K., Sandell, G., Hayashi, S., 
 Duncan, W.D., \& Griffin, M.J. 1990. B335 - Protostar or embedded 
 pre-main-sequence star? {\refit Mon.\ Not.\ Roy.\ Astron.\ Soc.\/} 243:330--335}
 
\ref{Chandler, C.J., Koerner, D.W., Sargent, A.I., \& Wood, D.O.S 1995. 
   Dust Emission from Protostars: The Disk and Envelope of HH 24 MMS. 
   {\refit Astrophys.\ J.\/} 449:L139--L142}

\ref{Chapman, S.J., Davies, J.R., Disney, M.J., Nelson, A.H., 
   Pongracic, H., Turner, J.A., Whitworth, A.P. 1992. The formation of binary 
   and multiple star systems. {\refit Nature\/} 359:207--210}

\ref{Chen, H., Grenfell, T. G., Myers, P. C., and Hughes, J. D. 1997. 
   Comparison 
   of Star Formation in Five Nearby Molecular Clouds. {\refit Astrophys.\ J.\/} 
   478:295--312 }

\ref{Chen, H., Myers, P. C., Ladd, E F., and Wood, D.O.S. 1995. Bolometric 
  temperature and young stars in the Taurus and Ophiuchus complexes.
  {\refit Astrophys.\ J.\/} 445:377--392 }

\ref{Chi\`eze, J.-P., and Pineau des For\^ets, G. 1987. The fragmentation of 
  molecular 
  clouds. II - Gravitational stability of low-mass molecular cloud cores. 
  {\refit Astron.\ Astrophys.\/} 183:98--108}

\ref{Chini, R., Kr\"ugel, E., Haslam, C.G.T., Kreysa, E., Lemke, R., 
   Reipurth, B., Sievers, A., \& Ward-Thompson, D. 1993. Discovery of a cold 
   and gravitationally unstable cloud fragment. 
   {\refit Astron.\ Astrophys.\/} 272:L5--L8 }

\ref{Chini, R., Reipurth, B., Sievers, A., Ward-Thompson, D., Haslam, C.G.T., 
   Kreysa, E., Lemke, R. 1997a. Cold dust around Herbig-Haro energy sources: 
   morphology and new protostellar candidates. 
   {\refit Astron.\ Astrophys.\/} 325:542--550}

\ref{Chini, R., Reipurth, B., Ward-Thompson, D., Bally, J., Nyman, L.A., 
   Sievers, A., \& Billawala, Y. 1997b. Dust Filaments and Star Formation in 
   OMC-2 and OMC-3. {\refit Astrophys.\ J.\/} 474:L135--L138}

\ref{Ciolek, G.E., \& K\"onigl, A. 1998. Dynamical Collapse of Nonrotating Magnetic 
   Molecular Cloud Cores: Evolution through Point-Mass Formation. 
   {\refit Astrophys.\ J.\/} 504:257--279}

\ref{Ciolek, G.E., \& Mouschovias, T.Ch. 1994. Ambipolar diffusion, 
   interstellar dust, and the formation of cloud cores and protostars. 
   III: Typical axisymmetric solutions. {\refit Astrophys.\ J.\/} 425:142--160}

\ref{Clemens, D. P., and Barvainis, R. 1988. A catalog of small, optically selected 
   molecular clouds - Optical, infrared, and millimeter properties.
  {\refit Astrophys.\ J.\ Sup.\/} 68:257--286}

\ref{Curiel, S., Raymond, J.C., Rodr\'\i guez, L.F., Cant\'o, J., 
   \& Moran, J.M. 1990. The exciting source of the bipolar outflow in L1448. 
   {\refit Astrophys.\ J.\ Lett.\/} 365:L85--L88} 

\ref{Curiel, S., Rodr\'\i guez, L.F., Moran, J.M., and Cant\'o, J. 1993. 
  The triple radio continuum source in Serpens - The birth of a 
  Herbig-Haro system? {\refit Astrophys.\ J.\/} 415:191--203}

\ref{Davidson, J.A. 1987. Low-Luminosity embedded sources and their environs.
  {\refit Astrophys.\ J.\/} 315:602--620}

\ref{Davis, C.J., and Eisl\"offel, J. 1995. Near-infrared imaging in H$_2$ 
  of molecular (CO) outflows from young stars. 
  {\refit Astron.\ Astrophys.\/}  300:851--869}

\ref{Dent, W.R.F., Matthews, H.E., and Walther, D.M. 1995. CO and shocked 
   H$_2$ in the highly collimated outflow from VLA 1623. 
   {\refit Mon.\ Not.\ Roy.\ Astron.\ Soc.\/} 277:193--209}

\ref{Dent, W.R.F., Matthews, H.E., and Ward-Thompson, D. 1998. The 
   submillimetre colour of young stellar objects. 
   {\refit Mon.\ Not.\ Roy.\ Astron.\ Soc.\/} 301:1049--1063}

%\ref{Dutrey, A., Guilloteau, S., and Gu\'elin, M. 1997.  
%{\refit Astron.\ Astrophys.\/} 317:L55--L58 }

\ref{Efstathiou, A., and Rowan-Robinson, M. 1991. Radiative transfer in 
   axisymmetric dust clouds. II - Models of rotating protostars. 
   {\refit Mon.\ Not.\ Roy.\ Astron.\ Soc.\/} 252:528--534}

\ref{Eiroa, C., Miranda, L.F., Anglada, G., Estalella, R., \& Torrelles, J.M. 
   1994. Herbig-Haro objects associated with extremely young sources in 
   L1527 and L1448. {\refit Astron.\ Astrophys.\/}  283:973--977}

\ref{Falgarone, E., \& Puget, J.-L. 1985. A model of clumped molecular clouds. 
   I - Hydrostatic structure of dense cores. 
   {\refit Astron.\ Astrophys.\/} 142:157--170}

\ref{Ferreira J., \& Pelletier G. 1995. Magnetized accretion-ejection structures. 
    III. Stellar and extragalactic jets as weakly dissipative disk outflows. 
   {\refit Astron.\ Astrophys.\/} 295:807--832}

\ref{Fiege, J.D., \& Henriksen, R.N. 1996. A global model of protostellar 
    bipolar outflow - I. {\refit Mon.\ Not.\ Roy.\ Astron.\ Soc.\/} 281:1038--1054}

\ref{Fletcher, A. B., and Stahler, S. W. 1994. The luminosity functions of embedded 
    stellar clusters. I: Method of solution and analytic results. 
   {\refit Astrophys.\ J.\/} 435:313--328 }

\ref{Foster, P. N., and Chevalier, R. A. 1993. Gravitational Collapse of an 
    Isothermal Sphere. {\refit Astrophys.\ J.\/} 416:303--311 (FC93) }

\ref{Fuller, G.A., Lada, E.A., Masson, C.R., Myers, P.C. 1995a. The 
    Infrared Nebula 
    and Outflow in Lynds 483. {\refit Astrophys.\ J.\/} 453:754--760}

\ref{Fuller, G.A., Lada, E.A., Masson, C.R., Myers, P.C. 1995b. 
    The Circumstellar 
    Molecular Core around L1551 IRS 5. {\refit Astrophys.\ J.\/} 454:862--871}

\ref{Gammie, C.F., and Ostriker, E.C. 1996. Can Nonlinear Hydromagnetic 
    Waves Support a Self-gravitating Cloud? 
    {\refit Astrophys.\ J.\/} 466:814--830 }

\ref{Gibb, A.G. 1999. A VLA search for embedded young stellar objects and
    protostellar candidates in L1630. 
    {\refit Mon.\ Not.\ Roy.\ Astron.\ Soc.\/} in press}

\ref{Gibb, A.G., \& Davis, C.J. 1998. The outflow from the class~0 protostar
    HH25MMS: methanol enhancement in a well-collimated outflow. 
    {\refit Mon.\ Not.\ Roy.\ Astron.\ Soc.\/}  298:644--656}

\ref{G\'omez, J.F., Curiel, S., Torrelles, J.M., Rodr\'\i guez, L.F., Anglada,
    G., and Girart, J.M. 1994. The molecular core and the powering source 
    of the bipolar molecular outflow in NGC 2264G. 
    {\refit Astrophys.\ J.\/} 436:749--753 }

\ref{G\'omez, M., Hartmann, L., Kenyon, S.J., and Hewett, R. 1993. 
    On the spatial distribution of pre-main-sequence stars in Taurus. 
    {\refit Astron.\ J.\/} 105:1927--1937 }

\ref{G\'omez, M., Whitney, B.A., and Wood, K. 1998.
    A Survey of Optical Jets and Herbig-Haro Objects in the rho Ophiuchi 
    Cloud Core. {\refit Astron.\ J.\/} 115:2018--2027 }

\ref{Greaves, J.S., and Holland, W.S. 1998. Twisted magnetic field lines 
    around protostars. {\refit Astron.\ Astrophys.\/} 333:L23--L26 }

\ref{Greaves, J.S., Holland, W.S., and Ward-Thompson, D. 1997. Submillimeter 
    Polarimetry of Class 0 Protostars: Constraints on Magnetized 
    Outflow Models. {\refit Astrophys.\ J.\/} 480:255--261}

\ref{Greene, T. P., and Lada C. J. 1996. Near-Infrared Spectra 
    and the Evolutionary 
    Status of Young Stellar Objects: Results of a 1.1-2.4~$\mu$m Survey. 
    {\refit Astron.\ J.\/} 112:2184--2221 }

%\ref{Greene, T. P., and Lada C. J. 1997. Near-Infrared Spectra of 
%Flat-Spectrum Protostars: Extremely Young Photospheres Revealed. 
%    {\refit Astron.\ J.\/} 114:2157--2165 }

\ref{Greene, T. P., Wilking B. A., Andr\'e P., Young E. T., and Lada C. J. 
     1994. Further mid-infrared study of the rho Ophiuchi cloud young stellar 
     population: Luminosities and masses of pre-main-sequence stars. 
     {\refit Astrophys.\ J.\/} 434:614--626 }

\ref{Gregersen, E. M., Evans II, N. J., Mardones, D., and Myers, P.C. 1999. 
Does Infall End Before the Class~I stage? {\refit Astrophys.\ J.\/} in press}

\ref{Gregersen, E. M., Evans II, N. J., Zhou, S., and Choi, M. 1997. 
     New Protostellar 
     Collapse Candidates: An HCO + Survey of the Class 0 Sources.
     {\refit Astrophys.\ J.\/} 484:256--276}

\ref{Grossman, E. N., Masson, C.R., Sargent, A.I., Scoville, N.Z., 
     Scott, S., and Woody, D.P 1987. A possible protostar near HH 7-11.
     {\refit Astrophys.\ J.\/} 320:356--363}

\ref{Gueth, F., Guilloteau, S., Dutrey, A., and Bachiller, R. 1997. Structure
   and kinematics of a protostar: mm-interferometry of L~1157. 
   {\refit Astron.\ Astrophys.\/} 323:943--952 }

\ref{G\"usten, R. 1994. Protostellar condensations. In {\refit The Cold 
 Universe\/}, eds. T. Montmerle, C.J. Lada, I.F. Mirabel, and J. Tr\^an Thanh 
 V\^an, (Gif-sur-Yvette: Editions Fronti\`eres), pp.\ 169--177 }

\ref{Henning, Th., and Launhardt, R. 1998. Millimetre study of star formation 
     in southern globules. {\refit Astron.\ Astrophys.\/} 338:223--242 }

\ref{Henning, Th., Michel, B., \& Stognienko, R. 1995. Dust opacities in dense 
     regions. {\refit Planet.\ Space Sci.\/} 43:1333--1343}

%\ref{Henriksen, R.N., \& Valls-Gabaud, D, 1994. 
%   {\refit Mon.\ Not.\ Roy.\ Astron.\ Soc.\/} 266:681--}

\ref{Henriksen, R.N., Andr\'e, P., and Bontemps, S. 1997. Time-dependent accretion
    and ejection implied by pre-stellar density profiles. 
    {\refit Astron.\ Astrophys.\/} 323:549--565 (HAB97) } 

\ref{Herbst, T.M., Beckwith, S, and Robberto, M. 1997. A New Molecular 
    Hydrogen Outflow in Serpens. 
    {\refit Astrophys.\ J.\ Lett.\/} 486:L59--L62}

\ref{Hildebrand, R.H. 1983. The determination of cloud masses and dust 
    characteristics from submillimetre thermal emission. 
    {\refit Quart.\ J.\ Roy.\ Astron.\ Soc.\/} 24:267--282 }

\ref{Hirano, N., Kameya, O., Kasuga, T., and Umemoto, T. 1992. Bipolar 
    outflow in B335 - The small-scale structure. 
    {\refit Astrophys.\ J.\ Lett.\/} 390:L85--L88}

\ref{Hodapp, K.-W., and Ladd, E. F. 1995. Bipolar Jets from Extremely Young
   Stars Observed in Molecular Hydrogen Emission. 
   {\refit Astrophys.\ J.\/} 453:715--720}

\ref{Hogerheijde, M. R. 1998. The molecular environment of low-mass protostars. 
     {\refit Ph.\ D.\ Dissertation,\ University\ of\ Leiden\/} 
     (Amsterdam: Thesis Publishers)}

\ref{Hogerheijde, M. R., van Dishoeck, E. F., Blake, G. A., and 
     van Langevelde, H. J. 1998. Envelope Structure on 700 AU Scales and the 
     Molecular Outflows of Low-Mass Young Stellar Objects. 
     {\refit Astrophys.\ J.\/} 502:315--336}

\ref{Hogerheijde, M. R., van Dishoeck, E. F., Salverda, J.M., and 
     Blake, G. A. 1999. Envelope Structure of deeply embedded young 
     stellar objects in the Serpens molecular cloud. 
     {\refit Astrophys.\ J.\/} in press}

\ref{Holland, W.S., Greaves, J.S., Ward-Thompson, D., and Andr\'e, P. 1996. 
    The magnetic field structure around protostars. Submillimetre polarimetry 
    of VLA~1623 and S106-IR/FIR. {\refit Astron.\ Astrophys.\/} 309:267--274 }

\ref{Hunter, T.R., Neugebauer, G., Benford, D.J., Matthews, K., Lis, D.C., 
    Serabyn, E., and Phillips, T.G. 1998. G34.24+0.13MM: A Deeply 
    Embedded Proto-B-Star. 
    {\refit Astrophys.\ J.\ Lett.\/} 493:L97--L100} 

\ref{Hurt, R. L., and Barsony, M. 1996. A Cluster of Class 0 Protostars 
    in Serpens: An IRAS HIRES Study. 
    {\refit Astrophys.\ J.\ Lett.\/} 460:L45--L48}

\ref{Hurt, R. L., Barsony, M., and Wootten, A. 1996. Potential Protostars 
   in Cloud Cores: H 2CO Observations of Serpens. 
   {\refit Astrophys.\ J.\/} 456:686--695}

\ref{Jenness, T., Scott, P. F., and Padman, R. 1995. Studies of embedded far-infrared
   sources in the vicinity of H$_2$0 masers - I. Observations. 
   {\refit Mon.\ Not.\ Roy.\ Astron.\ Soc.\/} 276:1024--1040}

\ref{Jessop, N., \& Ward-Thompson, D. 1999. Star Formation in molecular 
cloud cores. {\refit Mon.\ Not.\ Roy.\ Astron.\ Soc.\/} in prep.}

\ref{Johnstone, D., and Bally, J. 1999. JCMT/SCUBA sub-millimeter wavelength 
    imaging of the integral-shaped filament in Orion. 
    {\refit Astrophys.\ J.\ Lett.\/} 510:L49--L53}

\ref{Kenyon, S.J. and Hartmann, L 1995. Pre-Main-Sequence Evolution in the 
    Taurus-Auriga Molecular Cloud. {\refit Astrophys.\ J.\ Suppl.\/} 
    101:117--171}

\ref{Kenyon, S. J., Calvet, N., and Hartmann, L. 1993a. The embedded 
    young stars 
    in the Taurus-Auriga molecular cloud. I - Models for spectral energy 
    distributions. {\refit Astrophys.\ J.\/} 414:676--694}

\ref{Kenyon, S.J., Brown, D.I., Tout, C.A., and Berlind, P. 1998. 
    Optical Spectroscopy of Embedded Young Stars in the Taurus-Auriga 
    Molecular Cloud. {\refit Astron.\ J.\/} 115:2491--2591}

\ref{Kenyon, S.J., Whitney, B.A., Gomez, M., and Hartmann, L. 1993b. 
    The embedded young stars in the Taurus-Auriga molecular cloud. II - 
    Models for scattered light images. {\refit Astrophys.\ J.\/} 414:773--792}

\ref{Klessen, R.S., Burkert, A., and Bate, M.R. 1998. Fragmentation of 
    Molecular Clouds: The Initial Phase of a Stellar Cluster. 
    {\refit Astrophys.\ J.\ Lett.\/} 501:L205--L208}

\ref{Kroupa, P., Tout, C. A., and Gilmore, G. 1993. The distribution of low-mass 
    stars in the Galactic disc. 
    {\refit Mon.\ Not.\ Roy.\ Astron.\ Soc.\/} 262:545--587}

\ref{Lada, C.J. 1987. Star formation - From OB associations to protostars. 
    In {\refit Star Forming Regions Proc. IAU Symp. 115\/}, 
    eds.\ M. Peimbert and J. Jugaku (Dordrecht: Reidel), pp.\ 1--18. } 

\ref{Lada, C.J., and Fich, M. 1996. The Structure and Energetics of a 
    Highly Collimated Bipolar Outflow: NGC 2264G.  
    {\refit Astrophys.\ J.\/} 459:638--652}

\ref{Lada, C.J., and Wilking, B. 1984. The nature of the embedded population 
    in the Rho Ophiuchi dark cloud - Mid-infrared observations. 
    {\refit Astrophys.\ J.\/} 287:610--621}

\ref{Lada, C.J., Alves, J., and Lada, E. A. 1996. Near-Infrared Imaging 
    of Embedded Clusters: NGC 1333. {\refit Astron.\ J.\/} 111:1964--1976}

\ref{Ladd, E.F., and Hodapp, K.-W. 1997. A Double Outflow from a Deeply 
   Embedded Source in Cepheus. {\refit Astrophys.\ J.\/} 474:749--759}

\ref{Ladd, E.F., Adams, F.C., Casey, S., Davidson, J.A., 
   Fuller, G.A., Harper, D.A., Myers, P.C., \& Padman, R. 
   1991. Far-infrared and submillimeter wavelength observations of star-forming 
   dense cores. II - Images. {\refit Astrophys.\ J.\/} 382:555--569}

\ref{Ladd, E.F., Fuller, G.A., and Deane, J.R. 1998. C18O and C17O Observations 
   of Embedded Young Stars in the Taurus Molecular Cloud. I. 
   Integrated Intensities 
   and Column Densities. {\refit Astrophys.\ J.\/} 495:871--890}

\ref{Langer, W.D., Castets, A., and Lefloch, B. 1996. The IRAS~2 and 
   IRAS~4 Outflows and Star Formation in NGC 1333. 
   {\refit Astrophys.\ J.\ Lett.\/} 471:L111--L114}

\ref{Larson, R.B. 1969. Numerical calculations of the dynamics of a collapsing 
   proto-star. {\refit Mon.\ Not.\ Roy.\ Astron.\ Soc.\/} 145:271--295}

\ref{Larson, R.B. 1985. Cloud fragmentation and stellar masses. 
  {\refit Mon.\ Not.\ Roy.\ Astron.\ Soc.\/} 214:379--398}

\ref{Launhardt, R. 1996. Dust emission from star-forming regions. IV. 
  Dense cores 
  in the Orion B molecular cloud. {\refit Astron.\ Astrophys.\/} 312:569--584 }

\ref{Launhardt, R., Mezger, P.G., Haslam, C.G.T., Kreysa, E., Lemke, R., 
  Sievers, A., and Zylka, R. 1996. Dust emission from star-forming regions. 
  IV. Dense cores in the Orion B molecular cloud.
  {\refit Astron.\ Astrophys.\/} 312:569--584 }

\ref{Lay, O. P., Carlstrom, J. E., and Hills, R. E. 1995. NGC 1333 IRAS 4: 
   Further Multiplicity Revealed with the CSO-JCMT Interferometer.
   {\refit Astrophys.\ J.\ Lett.\/} 452:L73--L76}

\ref{Lee, C.W, and Myers, P.C. 1999. A catalogue of optically selected cores.
   {\refit Astrophys.\ J.\ Suppl.\/}  in press}

\ref{Lefloch, B., Castets, A., Cernicharo, J., and Loinard, L. 1998. 
   Widespread SiO emission in NGC1333. {\refit Astrophys.\ J.\ Lett.\/} 
   504:L109--L112}

\ref{Lefloch, B., Eisl\"offel, J., Lazareff, B. 1996. The remarkable 
   Class 0 source Cep~E. {\refit Astron.\ Astrophys.\/} 313:L17--L20 }

\ref{Lehtinen, K. 1997. Spectroscopic evidence of mass infall towards an 
   embedded infrared source in the globule DC~303.8-14.2.
   {\refit Astron.\ Astrophys.\/} 317:L5--L8 }

\ref{Leous, J.A., Feigelson, E.D., Andr\'e, P., \& Montmerle, T. 1991. 
   A rich cluster of radio stars in the Rho Ophiuchi cloud cores.
   {\refit Astrophys.\ J.\/}  379:683--688}

\ref{Li, Z.-Y. 1998. Formation and Collapse of Magnetized Spherical Molecular 
   Cloud Cores. {\refit Astrophys.\ J.\/}  493:230--246}

\ref{Li, Z.-Y., and Shu, F.H. 1996. Magnetized Singular Isothermal Toroids. 
   {\refit Astrophys.\ J.\/}  472:211--224}

\ref{Li, Z.-Y., and Shu, F.H. 1997. Self-similar Collapse of an 
   Isopedic Isothermal Disk. {\refit Astrophys.\ J.\/}  475:237--250}

\ref{Looney, L.W., Mundy, L.G., and Welch, W.J. 1999. Unveiling the Envelope 
   and Disk: A sub-arcsecond survey. {\refit Astrophys.\ J.\/}  }

\ref{Lucas, P. W., and Roche P. F. 1997. Butterfly star in Taurus: 
   structures of young stellar objects. 
   {\refit Mon.\ Not.\ Roy.\ Astron.\ Soc.\/} 286:895--919}

\ref{Mardones, D., Myers, P. C., Tafalla, M., Wilner, D. J., Bachiller, R.,
   Garay, G. 1997. A Search for Infall Motions toward Nearby Young Stellar 
   Objects. {\refit Astrophys.\ J.\/} 489:719--733}

\ref{Mauersberger, R., Wilson, T.L., Mezger, P.G., Gaume, R., Johnston, K.J. 1992. 
   The internal structure of molecular clouds. III - Evidence for molecular 
   depletion in the NGC~2024 condensations. 
   {\refit Astron.\ Astrophys.\/} 256:640--651}

\ref{McCaughrean, M.J., Rayner, J.T., \& Zinnecker, H. 1994. Discovery of a 
   molecular hydrogen jet near IC 348. {\refit Astrophys.\ J.\/} 436:L189--L192}

\ref{McKee, C.F. 1989. Photoionization-regulated star formation and the structure 
   of molecular clouds. {\refit Astrophys.\ J.\/} 345:782--801}

\ref{McLaughlin, D.E., and Pudritz, R.E. 1997. 
   Gravitational Collapse and Star Formation in Logotropic and 
   Nonisothermal Spheres {\refit Astrophys.\ J.\/} 476:750--765}

\ref{McMullin, J.P., Mundy, L.G., Wilking, B.A., Hezel, T., and Blake, G.A. 
   1994. Structure and chemistry in the northwestern condensation of the 
   Serpens molecular cloud core. {\refit Astrophys.\ J.\/} 424:222--236}

\ref{Men'shchikov, A.B. \& Henning, T. 1997. Radiation transfer in 
   circumstellar disks. {\refit Astron.\ Astrophys.\/} 318:879--907}

\ref{Menten, K.M., Serabyn, E., G\"usten, R., and Wilson, T.L. 1987. 
   Physical conditions in the IRAS 16293-2422 parent cloud. 
   {\refit Astron.\ Astrophys.\/} 177:L57--L60 }

\ref{Mezger, P.G., Sievers, A.W., Haslam, C.G.T., Kreysa, E., 
   Lemke, R., Mauersberger, R., \& Wilson, T.L. 1992. Dust emission from star 
   forming regions. II - The NGC 2024 cloud core: Revisited. 
   {\refit Astron.\ Astrophys.\/} 256:631--639}

\ref{Mizuno, A., Onishi, T., Hayashi, M. et al. 1994. Molecular cloud 
   condensation as a tracer of low-mass star formation. {\refit Nature} 
   368:719--721}

\ref{Molinari, S., Testi, L., Brand, J., Cesaroni, R., and Palla, F. 1998. 
   IRAS~23385$+$6053: A Prototype Massive Class 0 Object. 
   {\refit Astrophys.\ J.\ Lett.\/} 505:L39--L42}

\ref{Moreira, M. C., Yun J. L., V\'azquez, R., and Torrelles J. M. 1997. 
   Thermal Radio Sources in Bok Globules. {\refit Astron.\ J.\/} 113:1371--1374}

\ref{Moriarty-Schieven G.H., Wannier P.G., Keene J., Tamura M. 1994. 
   Circumprotostellar environments. 2: Envelopes, activity, and evolution.
   {\refit Astrophys.\ J.\/} 436:800--806}

\ref{Motte, F. 1998. Structure des coeurs denses proto-stellaires: 
  Etude en continuum millim\'etrique. 
  {\refit Ph.\ D.\ Dissertation,\ University\ of\ Paris\ XI\/} }

\ref{Motte, F., Andr\'e, P. et al. 1999. Density structure of 
   isolated protostellar envelopes: A millimeter continuum survey of
   Taurus infrared protostars. {\refit Astron.\ Astrophys.\/} in preparation}

\ref{Motte, F., Andr\'e, P., and Neri, R. 1998. The initial conditions of star  
  formation in the $\rho$ Ophiuchi main cloud: wide-field millimeter 
  continuum mapping. {\refit Astron.\ Astrophys.\/} 336:150--172}

\ref{Mouschovias, T.Ch. 1991. Single-stage fragmentation and a modern theory of 
  star formation. In {\refit The Physics of Star Formation and Early 
Stellar Evolution\/},  eds. C. J.Lada \& N. D. Kylafis, pp.\ 449-468}
 
\ref{Mouschovias, T.Ch. 1995. Role of magnetic fields in the early stages of star 
  formation. In {\refit The Physics of the Interstellar Medium and 
  Intergalactic Medium\/}, eds. A. Ferrara, C.F. Mc Kee, C. Heiles, 
  \& P.R. Shapiro (San Francisco: ASP), 80:184--217}

\ref{Mundy, L.G., Wootten, H.A., Wilking, B.A., Blake, G.A., 
   \& Sargent, A.I. 1992. IRAS 16293 - 2422 - A very young binary system? 
   {\refit Astrophys.\ J.\/} 385:306--313}

\ref{Mundy, L.G., McMullin, J.P., and Grossman, A.W. 1993. 
 Observations of circumstellar disks at centimeter wavelengths. 
 {\refit Icarus\/} 106:11--19} 

\ref{Myers, P.C. 1998. Cluster-forming Molecular Cloud Cores. 
  {\refit Astrophys.\ J.\ Lett.\/} 496:L109--L112}

\ref{Myers, P. C., and Benson, P. J. 1983. Dense cores in dark clouds. 
  II. NH$_3$ observations and star formation. 
  {\refit Astrophys.\ J.\/} 266:309--320}

\ref{Myers, P. C., and Fuller, G. A. 1993. Gravitational formation times 
  and stellar mass distributions for stars of mass 0.3-30~$M_\odot$. 
  {\refit Astrophys.\ J.\/} 402:635--642}

\ref{Myers, P.C., and Ladd, E.F. 1993. Bolometric temperatures of young stellar 
  objects. {\refit Astrophys.\ J.\ Lett.\/} 413:L47--L50}

\ref{Myers, P.C., Adams, F.C., Chen, H., and Schaff, E. 1998. Evolution of the 
  Bolometric Temperature and Luminosity of Young Stellar Objects. 
  {\refit Astrophys.\ J.\/} 492:703--726}

\ref{Myers, P.C., Bachiller, R., Caselli, P., Fuller, G.A., Mardones, D., 
  Tafalla, M., \& Wilner, D.J. 1995. Gravitational Infall in the Dense Cores 
  L1527 and L483. {\refit Astrophys.\ J.\ Lett.\/}  449:L65--L68}

\ref{Myers, P.C., Fuller, G.A., Mathieu, R.D., Beichman, C.A., Benson, P.J.,
 Schild, R.E., and Emerson, J.P. 1987. Near-infrared and optical observations 
 of IRAS sources in and near dense cores. {\refit Astrophys.\ J.\/} 319:340--357}

\ref{Myhill, E. A., and Kaula, W. M. 1992. Numerical models for the collapse 
 and fragmentation of centrally condensed molecular cloud cores.
 {\refit Astrophys.\ J.\/} 386:578--586}

\ref{Nakano, T. 1984. Contraction of magnetic interstellar clouds. 
   {\refit Fund.\ Cosm.\ Phys.\/} 9:139--231}

\ref{Nakano, T. 1998. Star Formation in Magnetic Clouds. 
   {\refit Astrophys.\ J.\/} 494:587--604}

\ref{Ohashi, N., Hayashi, M., Ho, P.T.P., and Momose, M. 1997. 
   Interferometric Imaging of IRAS 04368+2557 in the L1527 Molecular 
   Cloud Core: A Dynamically Infalling Envelope with Rotation. 
   {\refit Astrophys.\ J.\/} 475:211--223}

\ref{O'Linger, J., Wolf-Chase, G.A., Barsony, M., and Ward-Thompson, D. 1999. 
  L1448~IRS2: A HIRES-Identified Class~0 Protostar. 
  {\refit Astrophys.\ J.\/} in press }

\ref{Ossenkopf, V., and Henning, Th. 1994. Dust opacities for protostellar 
  cores. {\refit Astron.\ Astrophys.\/} 291:943--959 }

\ref{Ouyed, R., and Pudritz, R.E. 1997. Numerical Simulations of 
   Astrophysical Jets from Keplerian Disks. I. Stationary Models. 
   {\refit Astrophys.\ J.\/} 482:712--732}

\ref{Padoan, P., Nordlund, A., \& Jones, B.J.T. 1997. 
{\refit Mon.\ Not.\ Roy.\ Astron.\ Soc.\/} 288:145--152}

\ref{Palla, F., and Stahler, S.W. 1991. The evolution of intermediate-mass 
   protostars. I - Basic results. {\refit Astrophys.\ J.\/} 375:288--299}

\ref{Parker, N. D., Padman, R., \& Scott, P.F. 1991. Outflows in dark clouds - 
   Their role in protostellar evolution. 
   {\refit Mon.\ Not.\ Roy.\ Astron.\ Soc.\/} 252:442--461}

\ref{Persi, P., Ferrari-Toniolo, M., Marenzi, A.R., Anglada, G., Chini, R., 
   Kr\"ugel, E., and Sepulveda, I. 1994. Infrared images, 1.3 mm continuum 
   and ammonia line observations of IRAS~08076$-$3556. 
   {\refit Astron.\ Astrophys.\/} 282:233--239 }

\ref{Phelps, R., and Barsony, M. 1999. Herbig-Haro Objects in Serpens \& Ophiuchus. {\refit Astron.\ J.\/}, in preparation}

\ref{Pollack, J.B., Hollenbach, D., Beckwith, S., Simonelli, D.P., Roush, T., 
   Fong, W. 1994. Composition and radiative properties of grains in molecular 
   clouds and accretion disks. {\refit Astrophys.\ J.\/} 421:615--639}

\ref{Pravdo, S.H., Rodr\'\i guez, L.F., Curiel, S., Cant\'o, J., 
   Torrelles, J.M., Becker, R.H., and Sellgren, K. 1985. 
   Detection of radio continuum emission from Herbig-Haro objects 1 and 2 
   and from their central exciting source. 
   {\refit Astrophys.\ J.\ Lett.\/} 293:L35--L38}

\ref{Preibisch, Th., Ossenkopf, V., Yorke, H.W., \& Henning, Th. 1993. 
  The influence of ice-coated grains on protostellar spectra. 
  {\refit Astron.\ Astrophys.\/} 279:577--588}

\ref{Pringle, J.E. 1989. On the formation of binary stars.   
  {\refit Mon.\ Not.\ Roy.\ Astron.\ Soc.\/}  239:361--370}

\ref{Pudritz, R.E., Wilson, C.D., Carlstrom, J.E., Lay, O.P., Hills, R.E., 
  and Ward-Thompson, D. 1996. Accretions Disks Around Class~0 Protostars: 
  The Case of VLA~1623. {\refit Astrophys.\ J.\ Lett.\/} 470:L123--L126}

\ref{Reipurth, B., Chini, R., Kr\"ugel, E., Kreysa, E., and 
  Sievers, A. 1993. Cold dust around Herbig-Haro energy sources: a 1300~$\mu$m 
  survey. {\refit Astron.\ Astrophys.\/} 273:221--238}

\ref{Reipurth, B., Nyman, L.-A., and Chini, R. 1996. Protostellar candidates 
  in southern molecular clouds. {\refit Astron.\ Astrophys.\/} 314:258--264}

\ref{Richer, J.S., Hills, R.E., \& Padman, R. 1992. A fast CO jet in Orion B.  
  {\refit Mon.\ Not.\ Roy.\ Astron.\ Soc.\/}  254:525--538}

\ref{Richer, J.S., Padman, R., Ward-Thompson, D., Hills, R.E., 
  \& Harris, A.I. 1993. The molecular environment of S106~IR.
  {\refit Mon.\ Not.\ Roy.\ Astron.\ Soc.\/}  262:839--854}

\ref{Ristorcelli, I., Serra, G., Lamarre, J.M. et al. 1998. Discovery of a 
  Cold Extended Condensation in the Orion A Complex. 
  {\refit Astrophys.\ J.\/} 496:267--273}

\ref{Safier, P.N., McKee, C.F., \& Stahler, S.W. 1997. 
  Star Formation in Cold, Spherical, Magnetized Molecular Clouds. 
  {\refit Astrophys.\ J.\/} 485:660--679}

\ref{Sandell, G. 1994. Secondary calibrators at submillimeter wavelengths. 
  {\refit Mon.\ Not.\ Roy.\ Astron.\ Soc.\/} 262:839--854}

\ref{Sandell, G., and Weintraub, D.A. 1994. 
  A submillimeter protostar mear LkH-alpha 198. 
  {\refit Astron.\ Astrophys.\/} 292:L1--L4}

\ref{Sandell, G., Aspin, C., Duncan, W.D., Russell, A.P.G., 
  \& Robson, E.I. 1991. NGC 1333 IRAS 4 - A very young, low-luminosity 
   binary system.
  {\refit Astrophys.\ J.\/} 376:L17--L20}

\ref{Sandell, G., Knee, L.B.G., Aspin, C., Robson, I.E., \& Russell, A.P.G.
  1994. A molecular jet and bow shock in the low mass protostellar binary 
  NGC~1333-IRAS2. 
  {\refit Astron.\ Astrophys.\/} 285:L1--L4}

\ref{Sandell, G., Avery, L.W., Baas, F. et al. 1999. A jet-driven, extreme 
  high-velocity outflow powered by a cold, low-luminosity protostar near 
  NGC2023. {\refit Astrophys.\ J.\/} in press}

\ref{Saraceno, P., Andr\'e, P., Ceccarelli, C., Griffin, M., and Molinari, 
  S. 1996a. An evolutionary diagram for young stellar objects.
  {\refit Astron.\ Astrophys.\/} 309:827--839}

\ref{Saraceno, P., D'Antona, F., Palla, F., Griffin, M., and Tommasi, E. 1996b. 
  The luminosity-mm flux correlation of Class~I soures exciting outflows.
  In {\refit The Role of Dust in the Formation of Stars\/}, eds. H.U. K\"aufl 
  and R. Siebenmorgen (Berlin: Springer), pp.\ 59--62.} 

\ref{Scalo, J. 1998. The IMF Revisited: A Case for Variations. 
  In {\refit The Stellar Initial Mass Function\/}, eds. G. Gilmore 
  and D. Howell, {\refit A.S.P. Conf. Series\/}, 142:201-- }

\ref{Shu, F.H. 1977. Self-similar collapse of isothermal spheres and star formation.
  {\refit Astrophys.\ J.\/} 214:488--497}

\ref{Shu, F.H., Adams, F.C., and Lizano, S. 1987. Star formation in molecular 
   clouds - Observation and theory. {\refit Ann.\ Rev.\ Astron.\ Astrophys.\/} 
   25:23--81}

\ref{Shu, F., Najita, J., Galli, D., Ostriker, E., and Lizano S. 1993. 
  The collapse of clouds and the formation and evolution of stars and disks. 
  In {\refit Protostars \& Planets {I}{I}{I}\/}, eds.\ E. H. Levy and
  J. I. Lunine (Tucson: Univ.\ of Arizona Press), pp.\ 3--45.}

\ref{Shu, F.H., Najita, J., Ostriker, E., Wilkin, F., Ruden, S., and Lizano, S. 
   1994. Magnetocentrifugally driven flows from young stars and disks. 
   I: A generalized model. {\refit Astrophys.\ J.\/} 429:781--796}

\ref{Silk, J. 1995. A theory for the initial mass function. 
  {\refit Astrophys.\ J.\ Lett.\/} 438:L41--L44}

\ref{Sonnhalter, C., Preibisch, T., and Yorke, H.W. 1995. Frequency dependent 
  radiation transfer in protostellar disks. 
  {\refit Astron.\ Astrophys.\/} 299:545--556}

\ref{Stahler S.W. 1988. Deuterium and the stellar birthline. 
  {\refit Astrophys.\ J.\/} 332:804--825 }

\ref{Stahler, S.W., \& Walter, F.M. 1993. Pre-main-sequence evolution and the
  birth population. In {\refit Protostars \& Planets {I}{I}{I}\/}, 
  eds.\ E. H. Levy and J. I. Lunine (Tucson: Univ.\ of Arizona Press), 
  pp.\ 405--428}

\ref{Stanke, T., McCaughrean, M., and Zinnecker, H. 1998. First results of 
  an unbiased H$_2$ survey for protostellar jets in OrionA. 
  {\refit Astron.\ Astrophys.\/} 332:307--313}

\ref{Tafalla, M., Mardones, D., Myers, P.C., Caselli, P., Bachiller, R., 
  and Benson, P.J. 1998. L1544: A Starless Dense Core with Extended Inward
  Motions. {\refit Astrophys.\ J.\/} 504:900--914}

\ref{Tamura, M., Hayashi, S.S., Yamashita, T., Duncan, W.D., 
    and Hough, J.H. 1993. Magnetic field in a low-mass protostar disk --  
    Millimeter polarimetry of IRAS 16293-2422. 
    {\refit Astrophys.\ J.\ Lett.\/} 404:L21--L24}

\ref{Terebey, S., and Padgett, D.L. 1997. Millimeter interferometry of 
  Class~0 sources: Rotation and infall towards L1448N.  
    In {\refit Herbig-Haro Flows and the Birth of Low Mass Stars. 
    Proc. IAU Symp. 182\/}, 
    eds.\ B. Reipurth and C. Bertout (Dordrecht: Kluwer), pp.\ 507--514. }

\ref{Terebey, S., Chandler, C.J., \& Andr\'e, P. 1993. The contribution of 
  disks and envelopes to the millimeter continuum emission from very young 
  low-mass stars. {\refit Astrophys.\ J.\/} 414:759--772}

\ref{Terebey, S., Vogel S.N., Myers P.C. 1989. High-resolution CO 
  observations of young low-mass stars. 
  {\refit Astrophys.\ J.\/} 340:472--478}

\ref{Testi, L., and Sargent, A.I. 1998. Star Formation in Clusters: A Survey of 
  Compact Millimeter-Wave Sources in the Serpens Core. 
  {\refit Astrophys.\ J.\ Lett.\/} 508:L91--L94}

\ref{Tinney, C.G. 1993. The faintest stars - The luminosity and mass functions 
  at the bottom of the main sequence. {\refit Astrophys.\ J.\/} 414:279--301 }

\ref{Tinney, C.G. 1995. The Faintest Stars: The Luminosity and Mass Functions 
  at the Bottom of the Main Sequence -- Erratum. 
  {\refit Astrophys.\ J.\/} 445:1017}

\ref{Tomisaka, K. 1996. Accretion in Gravitationally Contracting Clouds. 
  {\refit Publ.\ Astron.\ Soc.\ Japan} 48:L97--L101 }

\ref{Umemoto, T., Iwata, T., Fukui, Y., Mikami, H., Yamamoto, S., Kameyama, O.,
  and Hirano, N. 1992. The outflow in the L1157 dark cloud - Evidence for 
  shock heating of the interacting gas.  
  {\refit Astrophys.\ J.\ Lett.\/} 392:L83--L86}

\ref{Velusamy, T., and Langer, W.D. 1998. Outflow-infall interactions as 
  a mechanism for terminating accretion in protostars. 
  {\refit Nature\/} 392:685--687}

\ref{Walker, C.K., Adams, F.C., and Lada, C.J. 1990. 1.3 millimeter 
  continuum observations of cold molecular cloud cores. 
  {\refit Astrophys.\ J.\/} 349:515--528}

\ref{Walker, C.K., Lada, C.J., Young, E.T., Maloney, P.R., \& Wilking, B.A. 
  1986. Spectroscopic evidence for infall around an extraordinary IRAS source 
  in Ophiuchus. {\refit Astrophys.\ J.\ Lett.\/} 309:L47--L51}

%\ref{Walker, C.K., Lada, C.J., Young, E.T., Margulis, M. 1988. An unusual 
%  outflow around IRAS~16293$-$2422. {\refit Astrophys.\ J.\/} 332:335--345}

\ref{Walker, C.K., Narayanan, G., \& Boss, A.P. 1994. Spectroscopic signatures 
  of infall in young protostellar systems. {\refit Astrophys.\ J.\/} 431:767--782}

\ref{Ward-Thompson, D. 1996. The Formation and Evolution of Low Mass 
  Protostars. {\refit Astrophys.\ Space Sci.\/} 239: 151--170}

\ref{Ward-Thompson, D., Eiroa, C., and Casali, M.M. 1995. 
   Confirmation of the driving source of the NGC 2264G bipolar outflow: 
   a Class 0 protostar. 
   {\refit Mon.\ Not.\ Roy.\ Astron.\ Soc.\/} 273:L25--L28}

\ref{Ward-Thompson, D., Motte, F., and Andr\'e, P. 1999. The initial conditions
  of isolated star formation. III: Millimetre continuum mapping of pre-stellar
  cores. {\refit Mon.\ Not.\ Roy.\ Astron.\ Soc.\/} in press (WMA99) }

\ref{Ward-Thompson, D., Buckley, H.D., Greaves, J.S., Holland, W.S., and 
  Andr\'e, P. 1996. Evidence for protostellar infall in NGC~1333-IRAS2. 
 {\refit Mon.\ Not.\ Roy.\ Astron.\ Soc.\/} 281:L53--L56}

\ref{Ward-Thompson, D., Chini, R., Krugel, E., Andr\'e, P., and Bontemps, S.
  1995. A submillimetre study of the Class 0 protostar HH24MMS.
  {\refit Mon.\ Not.\ Roy.\ Astron.\ Soc.\/} 274:1219--1224}

\ref{Ward-Thompson, D., Scott, P. F., Hills, R. E., and Andr\'e, P. 1994. 
  A submillimetre continuum survey of pre-protostellar cores. 
  {\refit Mon.\ Not.\ Roy.\ Astron.\ Soc.\/} 268:276--290 }

\ref{White, G.J., Casali, M.M., and Eiroa, C. 1995. 
   High resolution molecular line observations of the Serpens Nebula. 
  {\refit Astron.\ Astrophys.\/} 298:594--605}

\ref{Whitney, B.A., and Hartmann, L. 1993. Model scattering envelopes of young 
  stellar objects. II - Infalling envelopes. 
  {\refit Astrophys.\ J.\/} 402:605--622}

\ref{Whitworth, A., and Summers, D. 1985. Self-similar condensation of 
  spherically symmetric self-gravitating isothermal gas clouds. 
  {\refit Mon.\ Not.\ Roy.\ Astron.\ Soc.\/} 214:1--25 }

\ref{Whitworth, A.P., Bhattal, A.S., Francis, N., and Watkins, S.J. 1996. 
  Star formation and the singular isothermal sphere.  
  {\refit Mon.\ Not.\ Roy.\ Astron.\ Soc.\/} 283:1061--1070 }

\ref{Wiesemeyer, H. 1997. The spectral signature of accretion in low-mass 
  protostars. {\refit Ph.\ D.\ Dissertation,\ University\ of\ Bonn\/} }

\ref{Wiesemeyer, H., G\"usten, R., Wink, J.E., and Yorke, H.W. 1997. 
   High resolution studies of protostellar condensations in NGC 2024. 
  {\refit Astron.\ Astrophys.\/} 320:287--299}

\ref{Wiesemeyer, H., G\"usten, R., Cox, P., Zylka, R., and Wright, 
   M.C.H. 1998. The Pivotal Onset of Protostellar Collapse: ISO's View 
   and Complementary Observations. 
   In {\refit Star Formation with ISO\/}, eds.\ J.L. Yun 
   \& R. Liseau, {\refit A.S.P. Conf. Series\/}, 132:189--194}

\ref{Wilking, B.A., Lada, C.J., \& Young, E.T. 1989. IRAS observations of the 
  Rho Ophiuchi infrared cluster - Spectral energy distributions and 
  luminosity function. {\refit Astrophys.\ J.\/} 340:823--852 (WLY89) } 

\ref{Wilking, B.A., Schwartz, R.D., Fanetti, T.M., and Friel, E.D. 1997. 
  Herbig-Haro Objects in the $\rho$ Ophiuchi Cloud. 
  {\refit Publ.\ Astron.\ Soc.\ Pacific} 109:549--553 }

\ref{Wilner, D.J., Welch, W.J., and Forster, J.R. 1995. 
  Sub-Arcsecond Imaging of W3(OH) at 87.7 GHz.  
 {\refit Astrophys.\ J.\ Lett.\/} 449:L73--L76}

\ref{Wolf-Chase, G.A., Barsony, M., Wootten, H.A., Ward-Thompson, D., 
  Lowrance, P.J., Kastner, J.H., and McMullin, J.P. 1998. 
  The Protostellar Origin of a CS Outflow in S68N. 
 {\refit Astrophys.\ J.\ Lett.\/} 501:L193--L198}

\ref{Wood, D.O.S., Myers, P.C., \& Daugherty, D. A. 1994. IRAS images of 
  nearby dark clouds. {\refit Astrophys.\ J.\ Sup.\/} 95:457--501}

\ref{Wootten, A. 1989. The Duplicity of IRAS 16293-2422: A Protobinary Star? 
  {\refit Astrophys.\ J.\/} 337:858--864}

\ref{Yorke, H. W., Bodenheimer, P., and Laughlin, G. 1995. 
  The formation of protostellar disks. II: Disks around intermediate-mass stars 
  {\refit Astrophys.\ J.\/} 443:199--208}

\ref{Yu, K. C., Bally, J., and Devine, D. 1997. Shock-Excited H$_2$ Flows in 
  OMC-2 and OMC-3. {\refit Astrophys.\ J.\ Lett.\/} 485:L45--L48}

\ref{Yun, J. L., Moreira, M. C., Torrelles, J. M., Afonso, J. M., 
   and Santos, N. C. 1996. A Search for Radio Continuum Emission From Young 
   Stellar Objects in Bok Globules. {\refit Astron.\ J.\/} 111:841--845}

\ref{Zavagno, A., Molinari, S., Tommasi, E., Saraceno, P., and Griffin, M. 
   1997. Young Stellar Objects in Lynds 1641: a submillimetre continuum study.  
   {\refit Astron.\ Astrophys.\/} 325:685--692}

\ref{Zhou, S., Evans II, N.J., K\"ompe, C., \& Walmsley, C.M. 1993. 
   Evidence for protostellar collapse in B335. 
   {\refit Astrophys.\ J.\/} 404:232--246}

\ref{Zinnecker, H., Bastien, P., Arcoragi, J.P., \& Yorke, H.W. 1992. 
   Submillimeter dust continuum observations of three low luminosity 
   protostellar IRAS sources. {\refit Astron.\ Astrophys.\/} 265:726--732}

\ref{Zinnecker, H., McCaughrean, M.J., and Rayner, J.T. 1998. 
   A symmetrically pulsed jet of gas from an invisible protostar in Orion. 
   {\refit Nature} 394:862--865}

\vfill\eject
\null

\vskip .5in
\centerline{\bf FIGURE CAPTIONS}
\vskip .25in

%\caption{Figure 1.\capskip  H$_2$CO($2_{12}$-$1_{11}$) line spectra of the 
%pre-stellar core L1544 and the Class~0 protostar L1527 
%suggesting slow infall motions in the former and faster infall in the 
%latter (from Myers et al. 1996).}

%
\caption{Figure 1.\capskip Dust continuum images of  
L1544 at 90~$\mu$m (a) and 200~$\mu$m (b) from ISOPHOT, 
at 850~$\mu$m (c) from SCUBA, and at 1.3~mm (d) from IRAM 30~m. 
A polarization E-vector, perpendicular to the B field, 
is overlaid on the 850~$\mu$m image (c), as measured with the SCUBA polarimeter.
The observed morphology is consistent with a 
magnetically-supported core that has flattened along 
the direction of the mean magnetic field.}

\caption{Figure 2.\capskip Spectral energy distributions of the 
pre-stellar core L1544 and the Class~0 protostar IRAS~16293, along 
with grey-body fits (see text). The L1544 SED is based on ISOPHOT, 
JCMT, and IRAM data from Ward-Thompson et al. (1999a,b). The IRAS~16293 SED
is based on IRAS, ISO-LWS (Correia, Griffin et al. 1999 in prep. -- see also Ceccarelli et al. 1998), and JCMT data (Sandell 1994). Note that 
a simple grey-body model cannot account for the 25~$\mu$m emission of 
IRAS~16293 and that a two-component model is required (Correia et al. 1999).}

\caption{Figure 3.\capskip (a) (left) Radial intensity profile of L1689B at 
1.3~mm illustrating that pre-stellar cores have flat inner 
density profiles (from Andr\'e, Ward-Thompson, \& Motte 1996). 
For comparison, the dotted curve shows 
a spherical isothermal model with $\rho(r) \propto r^{-1.2}$ for $r < 4000$~AU  
and $\rho(r) \propto r^{-2}$ for $r \geq 4000$~AU.
The dash-dotted curve shows a model with $\rho(r) \propto r^{-2}$ such 
as a singular isothermal sphere (SIS). 

\noindent
~~~(b) (right) Typical density profiles expected for a
magnetically supported core undergoing ambipolar diffusion 
at various times increasing from $t_0$ to $t_6$ (from 
Ciolek \& Mouschovias 1994). The normalization values are 
$n_{n,c0} = 2.6 \times 10^3$~cm$^{-3}$ and $R_0 = 4.29$~pc.
Open circles mark the instantaneous radius  
of the uniform-density central region; starred circles mark the 
radius of the magnetically supercritical region 
(present only for $t \geq t_2$).}

\caption{Figure 4.\capskip Plot of lifetime vs. mean 
density for six core samples 
compared with $t \propto \rho^{-0.75}$ and 
$t \propto \rho^{-0.5}$ as predicted by ambipolar diffusion 
%in uniform magnetic field core 
models with different ionisation mechanisms 
(Jessop \& Ward-Thompson 1999).}

\caption{Figure 5.\capskip Dust continuum mosaic of the $\rho$~Oph cloud 
taken at 1.3~mm 
with the IRAM~30m telescope and the MPIfR bolometer array  
(Motte, Andr\'e, \&~Neri 1998). 
%The field is $\sim$~30'~$ \times$~15' in size, corresponding to 
%1.5~pc~$\times$~0.7~pc at $d = 160$~pc.   
} 

\caption{Figure 6.\capskip  Mass spectrum of  
the 59 pre-stellar fragments extracted 
from the $\rho$~Oph 1.3~mm continuum mosaic shown in  
Fig. 5 (from Motte et al. 1998). 
For comparison, dotted and 
long-dashed lines show power laws of the form 
$\Delta$N/$\Delta$M~$\propto$~M$^{-1.5}$ and 
$\Delta$N/$\Delta$M~$\propto$~M$^{-2.5}$, respectively.  
This pre-stellar mass spectrum is remarkable in that it resembles the shape 
of the IMF.}

\caption{Figure 7.\capskip 
%(a) $L_{bol}$--$T_{bol}$ diagram for YSOs embedded in the Taurus and 
%Ophiuchus clouds (from Chen et al. 1995).
(a) $L_{bol}$--$T_{bol}$ diagram for 14 well-documented YSOs along with 
three model evolutionary tracks for various (final) stellar masses and 
cloud temperatures (from Myers et al. 1998). Four times $t$~(Myr) since the
start of infall are indicated, 
at log~$t = -1.5$, $-1.0$, $-0.5$, and $0.0$.

\noindent
~~~(b) $M_{env}$--$L_{bol}$ diagram for a sample of Class~I 
(filled circles) and confirmed 
Class~0 sources (open circles) mainly in Ophiuchus, Perseus, and Orion 
(adapted from AM94 and Saraceno et al. 1996a). 
%Typical error bars are a factor of $\sim 2$ in $M_{env}$ and $\sim 50\% $ 
%in $L_{bol}$. Protostellar 
Evolutionary tracks, computed assuming 
protostars form from bounded condensations of finite initial masses 
and have $L_{bol} = GM_{\star}\dot{M}_{acc}/R_{\star} + L_{\star}$, where 
%$R_{\star}(M_{\star})$ is the protostellar radius and 
$L_{\star}$ is the PMS birthline luminosity (e.g. Stahler 1988), 
are shown.  \menv and $\dot{M}_{acc} = M_{env}/\tau$ 
(where $\tau = 10^5$~yr) have been assumed to decline exponentially with time 
%according to 
%$\maccm = \menvm/\tau = M_{cl}\, exp(-t/\tau)/\tau $, where $\tau = 10^5$~yr 
(see Bontemps et al. 1996a).
Small arrows are plotted on the tracks every $10^4$~yr, 
big arrows when 50\% and 90\% of the initial condensation has been accreted.
%For $M_{cl} = 2\, M_\odot $, an alternate (dash-dotted) track assuming a
%constant $\maccm = 10^{-5}\, \myr $ is also shown to illustrate the 
%difference in luminosity evolution. 
The dashed and dotted lines are two 
$M_{\star}$--$L_{bol}$ relations marking the conceptual border zone 
between the Class~0 ($M_{env} > M_\star $) 
and the Class~I ($M_{env} < M_\star $) stage; the dashed line has  
$M_\star \propto L_{bol}$ (cf. AWB93 and AM94) while 
the dotted line has $M_\star \propto L_{bol}^{0.6}$ as suggested by 
the accretion scenario adopted in the tracks. 
}

\caption{Figure 8.\capskip Normalized outflow momentum flux, 
$\fcom \, \rm c /\lbolm$, versus normalized envelope mass, 
$\menvm / \lbolm^{0.6} $, for a sample of 
Class~0 (open circles) and Class~I (filled circles) objects 
(from Bontemps et al. 1996a). 
$\fcom \, \rm c /\lbolm$ 
can be taken as an empirical tracer of the accretion rate $\dot{M}_{\rm acc}$, 
while $\menvm / \lbolm^{0.6} $ is an evolutionary indicator which decreases 
with time. 
This diagram should therefore mainly reflect the evolution of  
\macc during the protostellar phase. The solid curve shows 
the accretion rate history predicted by the simplified collapse model of 
Henriksen et al. (1997).}

\bye